\documentclass[10pt]{article}

\usepackage[backend=biber, natbib=true, doi=false,isbn=false,url=false,eprint=false]{biblatex} 
\usepackage{amsmath}
\usepackage{bm}
\usepackage{amssymb}
\usepackage{amsthm} 
\usepackage{dsfont}

\usepackage{subcaption}
\usepackage{graphicx}
\usepackage{color}
\usepackage{enumitem}

\usepackage{hyperref}
\usepackage{booktabs}

\newcommand{\de}{\,\mathrm{d}}
\usepackage{changes}

\theoremstyle{plain}
\newtheorem{theorem}{Theorem}

\theoremstyle{remark}
\newtheorem{remark}{Remark}
\usepackage{array}

\newcommand{\lla}{\left\langle}
\newcommand{\rra}{\right\rangle}

\theoremstyle{definition}
\newtheorem{example}{Example}[section]

\title{Ergodic Network Stochastic Differential Equations}

\author{Francesco Iafrate\footnote{University of Hamburg,
            Department of Mathematics,
            undestr. 55, 20146 Hamburg, Germany. Email: \texttt{francesco.iafrate@uni-hamburg.de}}
          \and
Stefano M. Iacus\footnote{
Harvard University, Institute for Quantitative Social Science, Cambridge, MA 02138, USA. Email: 
\texttt{siacus@iq.harvard.edu}} }

\begin{document}
	
	\maketitle
\begin{abstract}
We propose a novel framework for Network Stochastic Differential Equations (N-SDE), where each node in a network is governed by an SDE influenced by interactions with its neighbors. The evolution of each node is driven by the interplay of three key components: the node's intrinsic dynamics (\emph{momentum effect}), feedback from neighboring nodes (\emph{network effect}), and a \emph{stochastic volatility} term modeled by Brownian motion.
Our primary objective is to estimate the parameters of the N-SDE system from high-frequency discrete-time observations. The motivation behind this model lies in its ability to analyze high-dimensional time series by leveraging the inherent sparsity of the underlying network graph.
We consider two distinct scenarios: \textit{i)  known network structure}: the graph is fully specified, and we establish conditions under which the parameters can be identified, considering the linear growth of the parameter space with the number of edges.  
\textit{ii) unknown network structure}: the graph must be inferred from the data. For this, we develop an iterative procedure using adaptive Lasso, tailored to a specific subclass of N-SDE models.
In this work, we assume the network graph is oriented, paving the way for novel applications of SDEs in causal inference, enabling the study of cause-effect relationships in dynamic systems.
Through extensive simulation studies, we demonstrate the performance of our estimators across various graph topologies in high-dimensional settings. We also showcase the framework's applicability to real-world datasets, highlighting its potential for advancing the analysis of complex networked systems.
\end{abstract}

\noindent%
\vfill

\newpage

\section{Introduction}
The study of temporal data on networks has received considerable attention in recent years. In such models, the relationships between temporal variables are represented by a graph structure, allowing for the analysis of high-dimensional and interconnected systems. One notable example is the Network Autoregressive (NAR) model introduced in \cite{nar}, which leverages the network structure to handle ultra-high-dimensional time series. The NAR model is defined as:
$$Y_{it} =\theta_0 +\theta_1  \sum_{j=1}^d
\bar a_{ij} Y_{j(t-1)} +\theta_2 Y_{i(t-1)} +\epsilon_{it}, \quad i= 1, \ldots, d,$$
where $\epsilon_{it}$ is a Gaussian noise.
Here $\bar A$ denotes  the normalized adjacency matrix, i.e., 
$\bar A = (\bar a_{ij}) =  \text{diag}(N_1^{-1}, \ldots , N_d^{-1})A$, and $A = (a_{ij})$ the true adjacency matrix with elements $a_{ij}=1$ if there is a connection between nodes $i$ and $j$, and 0 otherwise. The quantities $N_i$ represent the number of neighbors of node $i$. In this model each component of $Y$ is represented as a node on the network.
The parameters $\theta_1$ and $\theta_2$ are termed respectively the \textit{momentum} (or node-effect) and \textit{network} (effect) parameter. The model can be rewritten as
$$ \mathbf{Y}_t = \mathcal T_0 +  \mathbf{Q}  \mathbf{Y}_{t-1} + \mathcal E_t,
$$
with $\mathcal T_0 = (\theta_0, \ldots, \theta_0)'$, $\mathcal E_t = (\epsilon_{1t}, \ldots,\epsilon_{dt})'$, and $\mathbf{Q} = \mathbf{Q}(\theta_1,\theta_2)$ defined as:
\begin{align}\label{eq:Q}
	\mathbf{Q} = \theta_1 \bar A +  \theta_2 I_{d\times d}.
\end{align}

In these models, the graph structure is assumed to be known but the dimension $d$ is allowed to grow at a rate which is compatible with the number of observations. 

The NAR model has been further extended in \cite{gnar} to account for higher-order neighbor interactions. These discrete-time models assume a known graph structure and allow the system's dimension 
$d$  to grow in relation to the sample size.

In continuous time, \cite{ou-graph1} introduced the $d$-dimensional Graph Ornstein-Uhlenbeck (GrOU) process, a 
$d$-dimensional system driven by L\'evy noise, defined as: 
$$\de Y_t =  - \mathbf{Q} Y_{t-}  \de_t +\de L_t$$
where $\mathbf{Q}_\theta$ is a $d\times d$ matrix with values in the positive cone $S^{++}$ defined as in equation~\eqref{eq:Q}, and $L_t$ is a $d$-dimensional L\'evy process.  The parameter $\theta = (\theta_1, \theta_2) \in R^2$ and conditions like $\theta_2 > 0$ such that $\theta_2 > |\theta_1|$, are required in order to guarantee ergodicity.
The parameters have the same interpretation: $\theta_1$ represents the \textit{momentum} effect and $\theta_2$ the  \textit{network} effect and are estimated via continuous time observations of $Y$. 
The same authors considered the inference under high frequency discrete-time observations  in  \cite{ou-graph2}. For both models they assume two cases: the matrix $A$ is fully known and specified, or unknown, in which case a LASSO approach is used to reconstruct it. In the GrOU framework though the dimension $d$ is not allowed to grow as in the NAR case.

\cite{koike2020-glasso} considered a 
$d$-dimensional semimartingale $Y = (Y_t)_{t\in [0,1]}$ with invertible variance-covariance matrix $\Sigma_Y = [Y,Y]_1$. The focus is to estimate the precision matrix  $\Theta_Y = \Sigma_y^{-1}$ under the asymptotic scheme $d\to\infty$. In this context the parametric structure of $Y$ is not relevant as the focus is on the elements of $\Theta_Y$, moreover the sparsity of the precision matrix is addressed through a weighted graphical lasso approach.

Motivated by these developments, we introduce a new framework for Network Stochastic Differential Equations (N-SDE). This framework generalizes the discrete-time NAR model and continuous-time GrOU processes by incorporating: \textit{i) non-linear} interactions between nodes through network effects and \textit{ii) stochastic volatility}, which propagates dynamically across the network.


\begin{align}\label{eq:nsde-intro}
	\de X_t^i &= \bigg( \underbrace{b_{ii}(X^i_t, \beta)}_{\mathsf{momentum \,\, effect}}  + \sum_{j \in N_{i}} \underbrace{ b_{ij}(X^i_t, X^j_t; \beta)}_{\mathsf{network \,\, effect}} \bigg) \de t +
	\underbrace{\sigma_{i}(X^i_t, \alpha)}_{\mathsf{node \,\, volatility}}\de W^i_t, 
\end{align}
$i = 1, 2, \ldots, d$.
In this model, each node on the network is represented by a stochastic differential equation. The evolution of node $i$ can be affected by its previous values as well as by nonlinear interactions with its neighbors $N_i$.  
By expanding on the decomposition proposed in \cite{nar}, the terms $b_{ii}$ represent a \emph{momentum effect}
whereas $ \sum_{j \in N_{i}}  b_{ij}$ measures \emph{network effect}. In our model, we further allow for a \emph{random volatility} term $\sigma_{ii}$ 
which determines volatility propagation across the network.

Our framework generalizes the work of \cite{NIPS2010} for the following linear system of SDEs: $$\de X_t = \sum_{j \in N_{i}} a_{ij} X_j^j \de t + \de W^i_t$$
for continuous time observations and repeated samples. In their framework the graph has a given structure represented by the elements $a_{ij}$ of the adjacent matrix and the objective is to estimate the minimal time horizon needed to fully estimate the network.
Similarly \cite{bishnoi2023graph} introduced the Brownian Graph Neural Network model defined by the following SDE:
$$
\de X_t^i = \left(X_t^i + \frac{F_{i}}{\gamma}\right)\de t + \sqrt{\frac{2 \kappa_B T}{\gamma}} \de W_t^i
$$
where $F_{i}$ is a function of both the `incoming to' and `outgoing from' edges for each node $X_t^i$, i.e., $F_i = \sum_{j : in} F_{ij} -  \sum_{j : out} F_{ij}$, $T$ is a fixed time horizon, and the rest are known parameters. The goal of this approach is to represent $F_{ij}$ as a graph neural network and estimate it using deep learning methods.
In a similar spirit \cite{bergna2023graph} introduced the Graph Neural SDE model that follows:
$$
\de X_t = f_\phi\left( X_t, t, \mathcal G\right)\de t + \sigma(X_t, t)\de W_t
$$
where $f_\phi$ is a neural network with weights $\phi$ and $\mathcal G$ represents the graph structure. The weights $\phi$ represent the quantity of interest.

Our model generalizes the above setups allowing for general non-linear SDE structures and parametric estimation under discrete-time sampling via quasi-likelihood estimation.
We analyze the case when the dimension of the parameter vector $(\alpha, \beta)$ is tied to the  growth of dimension $d$, in the setting when the graph structure is known providing the underlying sparsity pattern. 

As a second task, we are interested in recovering the structural information from multivariate time series, i.e. when the underlying graph structure is not known.
Topology and causal discovery in high-dimensional  time series via regularized estimation has been widely studied in the recent years, see for example in \cite{basu2015network},  \cite{JMLR:v11:songsiri10a}, \cite{songsiri2013sparse}.
In our N-SDE context we adopt adaptive regularized estimation techniques for ergodic diffusion processes, 
as developed in  \citep{de2012adaptive, de2021regularized, de2024pathwise}. 

In both cases the estimation is based on high frequency discrete-time observations from the model. 
Our approach is based on quasi-likelihood methods for parameter estimation, and we adopt the framework of \citep{yoshida1992estimation, yoshida2011polynomial}.

\paragraph{Open questions and main contributions.}

We now outline how our approach tackles some of the main gaps in the existing literature.

\begin{enumerate}

    \item[(i)] \textit{Restrictive linearity assumptions.}
    The linearity assumption in models like NAR or the Graphical OU process might be too restrictive in many real life applications. Stochastic volatility is also commonly observed in time phenomena.  Moreover it might be hard to verify what the true relation is. 
    
    \item[] \textit{Our contribution: non-linear effects and volatility}. Our framework accommodates general nonlinear effects under certain regularity assumptions. This flexibility might also allow for basis expansions (e.g., in $b_{ii}, b_{ij}$) for richer representations. The model also handles node-specific or state-dependent volatilities $\sigma_i(X)$, which can capture heteroskedastic behavior.

    \item[(ii)] \textit{Asymptotic results only.}
    Existing methods generally provide only asymptotic guarantees, offering limited insight into how the network size affects inference in finite samples. This leaves open questions about the scalability of those approaches as the graph grows.

    \item[] \textit{Our contribution: finite sample guarantees for growing networks.}
    In contrast to classical asymptotic results, our estimation theory leverages the graph structure to deliver non-asymptotic results, providing explicit error bounds such as 
    $\| \hat \theta_n - \theta_0 \|^2 = \mathcal O_p (|E| / n \Delta_n)$.
    Our results hold for growing networks, where both $d$ (the number of nodes) and
    $ |E| $, (the number of edges) scale with the sample size  $n$, provided that the network meets certain scaling conditions.

    \item[(iii)] \textit{Lack of directional relations.}
    Many real-world processes—particularly those involving causal or one-sided influences—cannot be adequately represented by symmetrical edges. Much of the literature focuses on undirected dependencies, potentially overlooking important directional effects.

    \item[] \textit{Our contribution: directed graphs.} 
    We incorporate oriented (directed) graphs to address this gap, enabling the model to learn one-way or asymmetric interactions among nodes. In many situations  flows of information or influence often travel in only one direction. By treating the adjacency matrix as directed, we can uncover and interpret these potentially causal pathways.

    \item[(iv)] \textit{Lack of interpretability in NN models.}
    Stochastic models based on (graph) neural networks can provide a powerful framework for deep learning and predicting high frequency time data.
    However, the nature of the inferred relations is unclear in such black-box models. 

    \item[] \textit{Our contribution: fully interpretable model.}
    Our model based approach provides a framework that allows for direct interpretation of the inferred relations. By directly modeling interactions in the drift and diffusion terms, our approach also helps practitioners incorporate prior knowledge (e.g., hypothesizing linear vs. nonlinear dependencies).

\end{enumerate}

The remainder of the paper is organized as follows. Section~\ref{sec:model} introduces the notation, the model, and the assumptions of ergodicity and  network scaling that ensure stability and statistical guarantees. 
Section ~\ref{sec:known-graph} analyzes a the N-SDE model from a non-asymptotic viewpoint, based on contrast regularity assumptions (\autoref{thm:err-bound}). 
Moreover, some explicit formulas in the linear drift, non-linear volatility case 
are derived in \autoref{sec:lin}.
Section~\ref{sec:lasso} considers the problem of graph recovery when the network structure is unknown. In \ref{thm:lasso-cons} we prove consistency of an adaptive Lasso procedure tailored to our network problem, in \autoref{thm:sel-cons} we provide a no-false-inclusion result for graph recovery.
Section~\ref{sec:sim} presents simulation studies to show the performance of the estimators under different graph structures and sample sizes. Finally, Section~\ref{sec:app} presents applications to real data, namely S\&P 500 stock prices.

\section{Network SDEs}\label{sec:model}

\paragraph{Model.}
Given a filtered probability space $(\Omega, \mathcal F, (\mathcal F_t)_{t \geq 0}, P)$ and an adapted 
$d$-dimensional Brownian motion $W =(W^1, \ldots, W^d)$, let $(X_t)_{t \geq 0}$ be the solution of the 
following system of stochastic differential equations: 
\begin{align}\label{eq:nsde}
	\de X_t^i &= \bigg( b_{ii}(X^i_t, \beta)  + \sum_{j \in N_{i}} b_{ij}(X^i_t, X^j_t; \beta) \bigg) \de t 
	+ 
	\sigma_{i}(X^i_t, \alpha)\de W^i_t  
\end{align}
$i=1, 2, \ldots  d$.
We further introduce the following notation to describe the graph: $N_i$ denotes the neighbours of node $i$ in a graph $G = (V, E)$, where $V = [d] := \{1, 2, \ldots, d\}$ is a known set of vertices, and $E$ is a fixed (i.e. non random) and known list of edges\footnote{In section~\ref{sec:lasso} we will consider $E$ deterministic but unknown.}. We write  $G_d= (V_d, E_d)$ to highlight the dependence on the dimension $d$.

The terms $b_{ij}$ and $\sigma_{i}$ denote the drift and diffusion functions of the model. They are known functions of unknown parameters $\theta = (\alpha, \beta)$.
We allow the dimension of the parameter space to grow with the size of the graph, i.e. $\pi_d^\alpha = \pi_\alpha (G_d)$, $\pi_d^\beta = \pi_\beta (G_d)$. We
denote the total number of parameters as $\pi_\theta(G_d) = \pi_d^\theta$.
The parameter space is denoted with $\Theta_d = \Theta_d^\alpha \times \Theta_d^\beta $, a compact subset of $\mathbb R^{\pi_d^\alpha + \pi_d^\beta}$. 
We denote with $\theta_0 \in \text{Int} \Theta_d$ the true value of the parameter.

We denote by $\bm{b} = (\bm{b}_{ij})$ the matrix whose elements are defined as:
\[
(\bm{b})_{ij} = \begin{cases}
	b_{ii} & i=j, \,\, i=1, \ldots, d \\
	b_{ij} & i=1, \ldots, d, \,\, j \in N_i \\
	0 &\text{ otherwise}
\end{cases}
\]
and set $\bm{\sigma} = \mathrm{diag}(\sigma_i, i = 1, \ldots, d)$. 
Model \eqref{eq:nsde} can  be rewritten in compact matrix form as 
\[
\de X_t =  (L_A \odot \bm{b}) \mathbf 1 \de t + 
 \bm{\sigma} \de W_t.
\]
where $L_A = I + A$, $A$ is the adjacency matrix, 
$\mathbf 1 =(1, \ldots, 1) \in \mathbb R^d$, $\odot$ is the Hadamard (element-wise) multiplication.
Denote with $\Sigma = \bm \sigma \bm \sigma ^\top$. 
Note that the functions $b_{ij}$ might include signs or normalization, this is why $L_A$ defined above does not correspond to the graph Laplacian, as one could expect in graph evolution models; for instance see Example \ref{ex:lin} for a reconciling case.

We consider discrete time observations from model \eqref{eq:nsde} under usual high-frequency asymptotics, i.e., 
the sample path of $X$ is observed at $n + 1$ equidistant discrete times $t_i^n$, such that $t_i^n-t_{i-1}^n=\Delta_n<\infty$ for $i=1,\ldots,n$ with $t_0^n=0$. 
We denote the discrete observations of the sample path of $X$ by $\mathbf{X}_n:=(X_{t_i^n})_{0\leq t_i\leq n}$, under the
following asymptotic scheme: $\Delta_n\longrightarrow 0$ as
$n\to \infty$,  $n\Delta_n \to 0$, in such a way that $n\Delta_n \geq n^{\epsilon_0}$, for some $\epsilon_0 >0$,  $n\Delta^2_n \to 0$. A sample path from model \eqref{eq:nsde} is shown in \autoref{fig:en-path}.

\paragraph{Assumptions.}
For \( l \geq 1 \) and \( m \geq 1 \), let \( f(x, \theta) \in C_\uparrow^{l,m}(\mathbb{R}^d \times \Theta, \mathbb{R}) \) be a space such that \( f(x, \theta) \) is continuously differentiable with respect to \( x \) up to order \( l \) for all \( \theta \), \( f(x, \theta) \) and all its \( x \)-derivatives up to order \( l \) are \( m \) times continuously differentiable with respect to \( \theta \) and \( f(x, \theta) \) and all derivatives are of polynomial growth in \( x \) uniformly in \( \theta \).

\noindent
In our setting $X$ is an ergodic diffusion process. Specifically, we assume the following set of conditions. 

\begin{enumerate}[label=($\mathbf{A\arabic*}$)]
    \item (Existence and uniqueness) \label{it:exist} There exists a constant $C$ such that
    $$\sup_{\beta\in\Theta_\beta}|b(x,\beta)-b(y,\beta)|+\sup_{\alpha\in\Theta_\alpha}||\sigma(x,\alpha)-\sigma(y,\alpha)|| \leq C|x-y|,\quad x,y\in \mathbb R^d.$$
       
    \item (Smoothness) \label{it:smooth} 
   $ b \in C_\uparrow^{0,4}(\mathbb{R}^d \times \Theta_\beta, \mathbb{R}^d)$
   and $ \sigma \in C_\uparrow^{2,4}(\mathbb{R}^d \times \Theta_\alpha, \mathbb{R}^d \otimes \mathbb{R}^{r})$. 
   
	\item \label{it:nondeg}(Non-degeneracy) There exists $\tau >0 $ such that  $\tau^{-1} \leq \Lambda_{\min} (\Sigma(x,\alpha))$, uniformly in $x$ and $\alpha$.
	
	\item \label{it:mix}(Mixing) There exists a positive constant $a$ such that
    $$\nu_X(u)\leq \frac{e^{-au}}{a},\quad u>0$$
    where
    $$\nu_X(u)=\sup_{t\geq 0}\sup_{\underset{B\in\sigma\{X_r:r\geq t+u\}}{A\in\sigma\{X_r:r\leq t\}}}|P(A\cap B)-P(A)P( B)|.$$
    
	  \item \label{it:mom}
     (Uniform boundedness) $\sup_t E[|X_t|^k]<\infty$ for all $k>0.$

	\item \label{it:ident}(Identifiability)
	$b(x, \beta) = b(x, \beta_0) \text{ for } \mu_{\theta_0} \, a.s. \text{ all }\, x \Rightarrow \alpha = \alpha_0$: 
	
	$\Sigma(x, \alpha) = \Sigma(x, \alpha_0) \text{ for } \mu_{\theta_0} \, a.s. \text{ all } \, x \Rightarrow \beta = \beta_0$ .
	
\end{enumerate}
\paragraph{Discussion of the assumptions and the ergodic property.} 
	Assumption \ref{it:nondeg} implies (D1)-(ii) of \cite{yoshida2011polynomial}.
    (see [D2] in \cite{yoshida2011polynomial}).
	The identifiability condition \ref{it:ident} is customary in the literature. It can be found for example in \cite{kessler1997estimation},  A6. 
    In particular, it implies that the random fields
	\[
	\mathbb{Y}(\alpha; \theta_0) = -\frac{1}{2} \int_{\mathbb{R}^d} \left\{ \mathrm{Tr} \left( {\Sigma}(x, \alpha)^{-1} {\Sigma}(x, \alpha_0) - {I}_d \right) + \log \frac{|{\Sigma}(x, \alpha)|}{|{\Sigma}(x, \alpha_0)|} \right\} \mu(dx).
	\]
	\[
	\mathbb{Y}(\beta; \theta_0) = -\frac{1}{2} \int_{\mathbb{R}^d} \langle \Sigma(x, \alpha_0)^{-1} ,(b(x, \beta) - b(x, \beta_0))^{\otimes 2} \rangle \mu(dx).
	\]
	are such that $\mathbb{Y} \neq 0$ for 
	$\theta \neq \theta_0$. This, and the fact that the model is defined on a compact set imply 
	[D3] and [D4] in \cite{yoshida2011polynomial}; on this point see  the remark in \cite[p. 462]{yoshida2011polynomial},  and \cite[p. 2894]{uchida2012adaptive}.

    The exponential mixing condition \ref{it:mix} implies that 
	$X$ is an \emph{ergodic} diffusion, namely that there exists a unique invariant probability measure $\mu=\mu_{\theta_0}$ such that
    \[ \frac{1}{T}\int_0^T g(X_t)\mathrm{d}t\overset{p}{\longrightarrow}\int_{\mathbb{R}^d}g(x)\mathrm{d}\mu
    \] 
    for any bounded measurable function $g:\mathbb{R}^d\to\mathbb{R}$.

    In order to verify assumptions \ref{it:mix} and \ref{it:mom} one can invoke the following results due to Pardoux and Veretennikov, which we recall here for the sake of the reader. 
    \begin{theorem}[\citet{pardoux2001poisson} - Prop. 3, \citet{veretennikov1988bounds} - main Theorem] \label{thm:ergo}
        Suppose that $\Sigma$ is bounded and there exist positive constants $\lambda_{-},\lambda_{+}$ and $\Lambda$ such that for all $\beta$
        
    \begin{equation}\label{eq:ergo1}
       0<\lambda_{-}\leq \langle \Sigma(x, \alpha)x/|x|, x/|x|\rangle\leq \lambda_{+},\quad \frac{\mathsf{Tr}(\Sigma(x, \alpha))}{d}\leq \Lambda 
    \end{equation}

    and, for all $\beta$,
    \begin{equation}\label{eq:ergo2}
        \langle b(x, \beta), x/|x|\rangle\leq - r|x|^a,\quad |x|\geq M_0, 
    \end{equation}
    with $M_0\geq 0, a\geq -1$ and $r>0$.
    Then the process is ergodic and the moment condition \ref{it:mom} holds. 
    If, in addition, $a \geq 1$, then $X$ satisfies the mixing condition \ref{it:mix}.
    \end{theorem}
    
We investigate in some detail the most straightforward specification of model \eqref{eq:nsde}, namely the linear case. 

\begin{example}[Linear effects] \label{ex:lin}
    Take $b_{ii} (x_i) = - \mu_i x_i, \, i \in [d]$, $b_{ij} = \beta_{ij} x_j, \,  (i,j) \in E$. 
    Denote by $\tau_{\max} (\cdot) $ the largest singular value of a matrix. 
    Denote by $B$ the \emph{weighted} adjacency matrix with weights $\beta_{ij}$, for a possibly directed graph $G$, and $\mu = (\mu_i, i \in [d])$.
    Then, by Cauchy-Schwarz inequality, the variational characterization of singular values and by Weyl's inequality,
    \begin{align*}
        \langle b(x) , x\rangle &= \langle -\mu I_d x + B x, x\rangle 
        \\ & \leq 
        (- \tau_{\max}(- \mu I_d) + \tau_{\max}(B) ) |x|^2
        \\ &= 
        -( \min_ i \mu_i - \tau_{\max} (B) ) |x|^2 .
    \end{align*}
    for $x \in \mathbb R^d$. Hence, \eqref{eq:ergo2} is satisfied if 
    \begin{equation} \label{eq:ergo-lin}
    \min_ i \mu_i  > \tau_{\max} (B).
    \end{equation}
    This, together with \eqref{eq:ergo1}, provides a sufficient condition for a linear, directed N-SDE model to be ergodic. 
    
    If, in addition, we assume that the weights are non-negative and symmetric, namely ${\beta_{ij} > 0} , \beta_{ij} = \beta_{ji}$ for all $(i,j) \in E $, by replacing $\tau_{\max} $ with the largest eigenvalue $\lambda_{\max}$ and by the Perron-Frobenius theorem, condition \eqref{eq:ergo-lin} is implied by 
    \begin{equation} \label{eq:ergo-lin2}
    \min_ i \mu_i  > \max_{i \in [d]} \sum_{j \in [d]} \beta_{ij}.
    \end{equation}
    For instance, in the case where momentum and network effects are constant across the network, namely $\mu_i = \mu_0, i \in [d]$, $\beta_{ij} = \beta_0, \, (i,j) \in E$, \eqref{eq:ergo-lin2} becomes $\mu_0 > \beta_0 \max_{i \in [d]} \mathsf{deg^-} (i)$, relating the coefficients' magnitude to the largest in-degree value.
\end{example}

More flexible non-linear class of models can be built as combinations of a dictionary of functions, say $\psi_j: \mathbb R^d \mapsto \mathbb R^d$, that is 
\[ 
b(x) = \sum_{j}\theta_j \psi_j(x).
\]

The following example shows a network dependent radial basis family satisfying ergodicity assumptions, adapting ideas from \cite{trottner2023concentration}, Example 1 in there.

\begin{example}\label{ex:radial‐decay}
Let $B_0 = \mathsf{diag} (\beta_{0, i}, i \in [d])$,
let $B_l = (\beta_{l, ij}, ~i, j \in [d]), l=0, 1,\ldots M)$ be parameter matrices, characterized as follows: for $l \geq 1$, $\beta_{l, ij} = 0$ if $(i, j) \notin E$, i.e. the parameter matrices for $l \geq 1$ are weighted adjacency matrices, while $B_0$ collects the momentum parameters.
For each basis index $l=1,\dots,M$ choose
$\alpha_l>0,\, q_l\in[-1,1]$ such that $q_1<q_2<\cdots<q_n$, and define 
the radial-type basis element
\[
\psi_0(x) = -x, \quad 
\psi_l(x)
\;=\;
(\alpha_l+\|x\|)^{-(q_l+1)}\,x,
\quad l \in [M], \, x \in \mathbb R^d
\]
A non-linear model for the drift can then be written as follows
\[
b(x)
\;=\;
\sum_{l=0}^M
B_l\,\psi_l(x),
\]
whose component $i$ reads
\[
b(x)_i \;=\; 
-\beta_{0, i} x_i + 
\sum_{l=1}^M
\sum_{j \in N_i}
\beta_{l, ij} \, x_{j}\;(\alpha_l+\|x\|)^{-(q_l+1)}.
\]
By combining the previous steps with the arguments in \cite{trottner2023concentration}, page 22, we get that a sufficient condition for ergodicity is 
given by 
\begin{equation} \label{eq:ergo-lin}
    \min_ i \beta_{0, i}  > \sum_{l=1}^M  \tau_{\max} (B_l).
\end{equation}

\end{example}

\bigskip

\section{Inference under Known Graph Structure}
\label{sec:known-graph}

In this section we analyze the properties of the quasi-likelihood estimator (defined below) under 
the assumption that the graph $G$ is known and $d$ fixed, but potentially very large. We use the notation $G_d$ to stress the dependence on the number of nodes $d$.

Our goal is to show that a N-SDE model can consistently handle large systems, as long as there is sufficient graph sparsity, and the observation period is long enough. 
We start by introducing the following assumption on the structure of the graph, describing the scenario in which we are working. Let $|G_d| := |V_d| + |E_d| = d + |E_d|$.

\begin{enumerate}[label=($\mathbf{ G\arabic*}$)]
	\item \label{it:par-scale} \emph{Network parametrization scaling}:
	For any $d\in \mathbb N$, 
	\[
	\frac{\pi_d}{|G_d|} \leq K		
	\]

	\item \label{it:net-scale} \emph{Graph scaling}:
	For any $d$, and for any $\epsilon$ there exists $n_0$ such that for any $n > n_0$
	\begin{equation*}
		\frac{|G_d|}{n \Delta_n} \leq   \epsilon .
	\end{equation*}
\end{enumerate}

\begin{remark}
	Conditions \ref{it:par-scale} and  \ref{it:net-scale} control the growth of the total number of parameters as the network dimension grows in terms of the number of observations $n$. It amounts to saying that each function is allowed to have an approximately constant number of parameters and the number of edges should be of the same order of the number of parameters of the model.
\end{remark}

\begin{example}
    Suppose $\pi^\alpha_d = 1$ for all $d$. For a linear effects model in Example \autoref{ex:lin}, \ref{it:par-scale} is satisfied with $K=2$, since $\pi_d = d + |E_d| + 1$. 
    For a dictionary of functions as in Example \ref{ex:radial‐decay},   \ref{it:par-scale} is satisfied with $K = M+2$.
\end{example}
 

We assume we observe data generated by model \eqref{eq:nsde}, where the neighborhoods $N_i$ are known. Our workhorse is the quasi-likelihood function for the parameter of interest $(\alpha, \beta)$, defined as 
\begin{align}\label{eq:ql-cond}
	\ell_n(\alpha, \beta) &= \\
	&\sum_{i=1}^{n}  \left\{ \frac{1}{2\Delta_n} \langle C^{-1}_{i-1}(\alpha) ,(\Delta X_{t_i} - b_{A, i-1}(\beta)  ) ^{\otimes 2} \rangle + \log \det C_{i-1}(\alpha) \right\}
	\notag
\end{align}
where $\Delta X_{t_i} = X_{t_i} - X_{t_{i-1}}$, $C_{i}(\alpha)= (\bm{\sigma} \bm{\sigma}^\top)(X_{t_i}; \alpha)$,
and $b_{A, i-1} = (L_A \odot \bm{b}(X_{t_{i-1}}, \beta)) \mathbf 1$.
The quasi-likelihood estimator 
\begin{equation}\label{eq:qmle-cond}
	\hat \theta_{n,d} = (\hat  \alpha_{n,d}, \hat \beta_{n,d}) \in \arg \min_{\alpha, \beta} \ell_n(\alpha, \beta).
\end{equation}

We write $\hat \theta_{n,d}$ so to stress the dependence of the estimator on both the sample size
$n$ and the dimension of the network $d$. We may omit subscripts for ease of read.
Throughout this section, $\hat \alpha$ denotes the estimator \eqref{eq:qmle-cond}. 

Denote with 
\[
\Gamma_n = 
\begin{pmatrix}
	\frac{1}{\sqrt n} \mathbf I_{\pi^\alpha} & 0 \\
	0 & \frac{1}{\sqrt{ n \Delta_n} } \mathbf I_{\pi^\beta} 
\end{pmatrix}
\qquad 
\]
the block matrix of the estimator rates and its graph-size scaled version.

In order to state our forthcoming result about a non-asymptotic error bound for estimation on a graph, we introduce regularity conditions on the contrast. Denote by $\partial_\theta \ell_n$
and  $\partial^2_{\theta, \theta} \ell_n$ the gradient and Hessian matrix of $\ell_n$, respectively, and by $\partial_\theta \overline{\ell_n} = \Gamma_n \partial_\theta \ell_n$, 
$\partial^2_{\theta, \theta} \overline{\ell_n} = \Gamma_n
\partial^2_{\theta, \theta} \ell_n \Gamma_n$ their scaled version.

\begin{enumerate}[]
	\item[\bf{C}$(r)$]
	\emph{Regular contrast}: 
	The functions $\ell_n$,  $\partial_\theta \ell_n$,  $\partial_\theta^2\ell_n$ can be extended continuously
	to the boundary of $\Theta$ and there exist square-integrable random variables $\xi_n$ with $\mathbb E\xi_n^2 \leq J$, and $\mu > 0$ s.t. 
	\[
	(i) \quad \max_{i \in [\pi_d]} \sup_{\theta:| \theta_0 - \theta| \leq r}|
	\partial_{\theta_i} \overline{\ell_n} | \leq \xi_n,
	\qquad 
	\]
	\[
	(ii) \quad \inf_{\theta:|\hat \theta_n - \theta| \leq r}  v^\top \partial^2 \overline{\ell_n}(\theta) v > \mu |v|^2 \qquad \forall v \in \mathbb R^{\pi_d},
	\]
	for all $n$, $P_{\theta_0} \, a.s.$.
\end{enumerate}

Assumption $C-(ii)$ is a fairly standard eigenvalue condition on the Hessian of the negative quasi-likelihood, and can be seen as  a finite sample  identifiability condition (compare with, e.g., \cite{ciolek2022lasso}, Assumption $\mathcal A$ (c)).
Condition $C-(i)$ relates to the regularity of the drift and diffusion functions $\bm b$ and $\bm \sigma$. The bounding variables $\xi_n$ can be characterized in terms of the polynomial growth condition of such terms.


The next theorem shows how the $\ell_2$-error of the estimator can be controlled with high probability by quantities related to the regularity of the model, the edge parametrization and the graph scaling. 

\begin{theorem}\label{thm:err-bound}
	Suppose that Assumptions \ref{it:exist} - \ref{it:ident}, \ref{it:par-scale} - \ref{it:net-scale} and $\mathbf C(r/n\Delta_n)$, hold true, for some $r>0$. Then, for every $\epsilon > 0, d>0$, there is $n_0$ such that for $n > n_0$ we have
	\begin{equation}\label{eq:err-bound}
		|\hat \theta_n - \theta_0|^2 \leq \frac{4 \xi^2_n}{\mu^2} K \epsilon.
	\end{equation}
	with probability at least $1 - C_L/r^L$, for some $L>0, C_L > 0$, not depending on $n$.
\end{theorem}

\begin{remark}
    In the preceding theorem, the parameter $r>0$ serves as a tuning parameter for those inequalities by controlling the finite-sample regularity of the contrast in a neighborhood of the maximum likelihood estimator.  For a fixed $n$, attaining a higher probability level forces assumption $\mathbf C\bigl(r/n\Delta_n\bigr)$ to hold on a wider neighborhood around the estimator.  Conversely, as $n$ increases, the regularity requirement becomes progressively less restrictive.

\end{remark}
\begin{remark}
	The proof of the theorem relies on deriving an error bound on the estimator depending on the number of parameters. In general, given an estimator 
	$\hat \theta_n$ the theorem could be proved under the following modified assumption: 
\end{remark}
\begin{enumerate}[label=($\mathbf{C\arabic*}'$)]
	\item
	\emph{Estimator scaling}:
	\[
	\sup_n \mathbb E | \Gamma_n^{-1} (\hat \theta - \theta_0)|^2 \lesssim \pi_d.
	\]
\end{enumerate}

\begin{remark}
	
	In \cite{yoshida2021simplified} general simplified conditions are given for an estimator to satisfy, for any $d$,
	\[
	\sup_n \mathbb E | \Gamma_n^{-1} (\hat \theta - \theta_0)|^p < \infty.
	\]
	Then, by the moment convergence in  Th. 3.5,
	one has that
	\[
	\mathbb E | \Gamma_n^{-1} (\hat \theta - \theta_0)|^2 \to \mathbb E |\Delta|^2,
	\]
	where $\Delta \sim \mathcal N(0, \mathcal I(\theta_0)^{-1}$, and then $\mathbb E |\Delta|^2 = \rm{tr} I(\theta_0)^{-1} \leq \pi_d | I(\theta_0)^{-1} |$. 
	
\end{remark}

\subsection{Linear N-SDE estimator}\label{sec:lin}
We turn our attention to the simple but important case of a N-SDE model with a linear drift. In this model the \emph{network effect} is given by a linear combinations of the parents of the node. We still allow for a nonlinear diagonal diffusion term and directed edges. We remark that, for readability, we present the results for the linear case; however, they can be readily extended to linear combinations of univariate basis functions. 

Suppose the drift functions take the form
\begin{equation}\label{eq:lin-drift}
	b_{ii}(X^i, \beta) = \beta_{0i} -\beta_{ii} X^i  \qquad b_{ij}(X^i, X^j; \beta) = \sum_{j \in N_i} \beta_{ij} X^{j}
\end{equation}
and that the diffusion matrix is diagonal $\bm \sigma = \mathsf{diag}(\sigma_j(x, \alpha), \, j \in [d])$.

Following \cite{uchida2012adaptive}, it is possible to use the following adaptive estimation procedure
\[
\hat \alpha_n  \in \arg \min_\alpha \mathcal U_n(\alpha) \qquad 
\hat \beta_n  \in  \arg \min_\beta \mathcal V_n(\hat \alpha_n, \beta)
\]
where
\begin{align}
	\mathcal U_n(\alpha) &= \frac{1}{\Delta_n} \sum_{i=1}^{n} \langle C^{-1}_{i-1}(\alpha) , \Delta X_{t_i}^{\otimes 2} \rangle + \log \det C_{i-1}(\alpha)
	\\
	\mathcal V_n(\alpha, \beta) &= \frac{1}{\Delta_n} \sum_{i=1}^{n} \langle C^{-1}_{i-1}(\alpha) , (\Delta X_{t_i} - \Delta_n b_{A, i-i}(\beta)^{\otimes 2} \rangle \label{eq:v-contrast}
\end{align}

We focus on the explicit form of $\hat \beta_n$ under the linearity assumption. 
In the case of diagonal noise, \eqref{eq:v-contrast} can be rewritten as
\begin{align*}
	\mathcal V_n(\hat \alpha_n, \beta) = 
	\frac{1}{2\Delta_n} \sum_{i=1}^n \sum_{j=1}^d 
	\frac{1}{\sigma^2_{j, t_{i-1}}(\hat\alpha_n)}
	\left[ \Delta X_{t_i}^j - \Delta_n \left(\beta_{0j} - \sum_{k \in N_j \cup \{j\}} \beta_{jk} X^k_{t_{i-1}}\right)\right]^2
\end{align*}
where $\bar N_j = N_j \cup \{j\} $. Let $\hat \sigma = \sigma(\hat \alpha_n)$. The score can be computed as
\begin{align*}
	\partial_{\beta_{jl}} \mathcal V_n= 
	-\sum_{i=1}^n 
	\frac{X^l_{t_{i-1}}}{\hat\sigma^2_{j, t_{i-1}}}
	\left[ \Delta X_{t_i}^j - \Delta_n \left(\beta_{0j} - \sum_{k \in N_j \cup \{j\}} \beta_{jk} X^k_{t_{i-1}}\right)\right].
\end{align*}
for $j \in [d], l \in [\bar N_j]$ (excluding the intercepts). In the case where the model has no intercepts, i.e $\beta_{0j} = 0 \, \forall j$, the estimating equations take the form
\begin{equation}\label{eq:lin-eq}
	\sum_{k \in \bar N_j} \beta_{jk} \sum_{i=1}^n \frac{X_{t_{i-i}}^k X_{t_{i-i}}^l}{\hat \sigma^2_{j, t_{i-1}}}
	= 
	\frac{1}{\Delta_n}
	\sum_{i=1}^n \frac{\Delta X_{t_i}^j X_{t_{i-i}}^l}{\hat \sigma^2_{j, t_{i-1}}}, 
	\quad j \in [d], l \in [\bar N_j].
\end{equation}
From a statistical point of view, each neighborhood behaves as a small $| \bar N_j|$-dimensional VAR model and the estimates of the parameters in a neighborhood only depend on the neighbours (but the estimators are not independent). In particular, denote with
$
\beta^{\bar N_j} = (\beta_j,\, j \in [\bar N_j])
$
the sub vector of parameters related to neighborhood $N_j$,
with 
$
\mathbf X_n = (X_{t_i}^j, i \in 0, \ldots, n-1, \, j \in [d])
$
the data matrix and let
$
\Delta \mathbf X_n = (X_{t_i}^j - X_{t_{i-1}}^j, i \in 1, \ldots, n, \, j \in [d])
$.
Similarly let
$
\mathbf X_n^{\bar N_j}
$
the columns of $\mathbf X_n$ corresponding to $\bar N_j$. Let 
$
\hat \sigma_{j,n} = (\hat \sigma_{t_{i}}(\hat \alpha_n), i =0 , \ldots, n-1)
$.
We write 
$
(\mathbf X^{\bar N_j})^{\otimes 2}_n =
(X_{t_i}^{\otimes 2}, i=0, \ldots, n-1)
$.
Let $\langle Y \rangle $ be  the matrix defined by
$
\langle Y \rangle_{ij} = n^{-1} \sum_{k=1}^{n} Y_{ij,k} \,  i \in [d_1], j\in [d_2]
$
for $Y \in \mathbb R^{d_1 \times d_2 \times n}$ (possibly a vector).

With this notation, \eqref{eq:lin-eq} can be rewritten as 
\begin{equation}
	\left \langle 
	\frac{
		(\mathbf X^{\bar N_j})^{\otimes 2}_n
	}{ \hat \sigma_{j,n}^2 } 
	\right \rangle
	\beta^{\bar N_j} =
	\frac{1}{\Delta_n} \left \langle 
	\frac{\Delta \mathbf X_n^{j}  \mathbf X_n^{\bar N_j} }{ \hat \sigma_{j,n}^2} 
	\right \rangle
\end{equation}
where the above division is meant in a vectorized sense. The estimator  $\hat \beta^{\bar N_j}$ can then be computed as 
\begin{equation}\label{eq:lin-est}
	\hat \beta^{\bar N_j} =
	\frac{1}{\Delta_n}
	\left \langle 
	\frac{(\mathbf X_n^{\bar N_j})^{\otimes 2}}{ \hat \sigma_{j,n}^2 } 
	\right \rangle^{-1}
	\left \langle 
	\frac{\Delta \mathbf X_n^{j}  \mathbf X_n^{\bar N_j} }{ \hat \sigma_{j,n}^2} 
	\right \rangle.
\end{equation}

The result above could be generalized to a linear drift model with diagonal  diffusion term of the form
\[
\sigma_j(x^{[\bar 
	N_j]}, \alpha) = \sqrt{\alpha_j + (x^j)^2 + f(x^{[N_j]})},
\]
i.e. with a neighborhood-dependent volatility term, with $f$ bounded and non-negative.

\section{Adaptive Lasso estimation of the graph structure}\label{sec:lasso}

We now consider the case where the adjacency matrix $A$ is not known. The goal is to recover the graph structure from the data using a regularization technique.

For this aim we need to slightly modify the setup as follows. We introduce auxiliary parameters $w$ that play the role of edge weights. Formally, we augment the parameter space 
as $(\theta, w) = (\alpha, \beta, w)$ with $w = \mathrm {vec} (w_{ij}, 1\leq i,j \leq d, \, j \neq i)$.
The $w$ parameters vary within the compact domain $\Theta_w \in \mathbb R^{d(d-1)}$. For ease of notation, we identify the entries of the vector $w$ with the extra-diagonal elements of a matrix -- that is we still write $w_{ij}$ for the weight corresponding to edge $(i,j)$. The true value $w_0 \in \text{Int} \Theta_w$ is such that 
$w_{0, ij} \neq 0$ if $A_{ij}=1$.
We recast model \eqref{eq:nsde} in the following 
form
\begin{align}\label{eq:nsde-w}
	\de X_t^i &= \bigg( b_{ii}(X^i_t, \beta)  + \sum_{j = 1, j \neq i}^d w_{ij}  b_{ij}'(X^i_t, X^j_t; \beta) \bigg) \de t 
	+ \sigma_{i}(X^i_t, \alpha)\de W^i_t,
\end{align}
so that the adjacency matrix $A$ of the graph is modeled as $A_{ij} = \mathds 1(w_{ij} \neq 0)$. In this formulation the edge pattern can be recovered by applying a LASSO-type regularization to the weights $w$. 

We denote by $ b_{ij}(x, y; \beta, w) = w_{ij}  b_{ij}'(x, y; \beta)$, for $i \neq j$
in the same spirit of model \eqref{eq:nsde}.
Denote with $(\theta_0, w_0)$ the true parameter value.

In order to ensure model identifiability we introduce the following extension to condition \ref{it:ident}:
\begin{enumerate}
	\item[\ref{it:ident}$'$]$b_{ij}(x, y; \beta)$ and $\sigma_i$ satisfy assumption \ref{it:ident} for all  $ i,j$ and, for any $\beta \in \Theta_\beta$
	\[
	b_{ij}(\cdot, \cdot; \beta, w) =0 \,\, \forall x , y \,\, \Leftrightarrow w_{ij} = 0, \quad i \neq j.
	\]
\end{enumerate}
This assumption ensures that any multiplicative constant in the model is modeled by the $w$ parameters.  

We propose a two-step procedure to estimate the graph and the parameters.  
In the first step we obtain an initial, 
non-regularized, estimate for the both 
the diffusion and drift parameters based on a consistent estimator. We focus on quasi-likelihood theory, even though this approach could be generalized to any consistent estimator, see e.g. \cite{de2021regularized}. We then 
build a penalized estimator of \emph{least squares approximation} (LSA) type. Such estimation strategy has been thoroughly investigated in the statistical literature (e.g., \cite{zou2006adaptive}, \cite{wang2007unified}) and was specifically applied to diffusion processes for lasso estimation by \cite{de2012adaptive}. Extensions incorporating non-convex penalties and $\ell_1$–$\ell_2$ (elastic net) regularization are discussed in \cite{suzuki2020penalized}, \cite{de2021regularized}, and \cite{de2024adaptive}.


Denote with 
$(\tilde  \theta_n, \tilde  w_n)$ the quasi-likelihood estimator of $(  \theta,   w)$ given by
\begin{equation}\label{eq:ql-est}
	(\tilde  \theta_n, \tilde  w_n) \in \arg \min_{\theta, w} \mathcal \ell_n(\theta, w) 
\end{equation}
where $\ell_n$ denotes the quasi-likelihood function \eqref{eq:ql-cond} computed with respect to the augmented model \eqref{eq:nsde-w}. 

In the following let $H_n$ be a data-dependent square information matrix of size $\pi^\alpha + \pi^\beta + d(d-1)$ matrix; let ${\mathcal  H}_n = \Gamma_n H_n \Gamma_n$ the corresponding scaled information matrix, 
where the scaling matrix now takes into account the additional $w$ parameters and is given by
\[
\Gamma_n = 
\begin{pmatrix}
	\frac{1}{\sqrt n} \mathbf I_{\pi^\alpha} & 0 \\
	0 & \frac{1}{\sqrt{ n \Delta_n} } \mathbf I_{\pi^\beta + d(d-1)}.
\end{pmatrix}
\]

Under standard regularity assumptions it has been established that the quasi-likelihood estimator above is $\Gamma_n$-consistent; moreover the empirical information based on the quasi-likelihood is uniformly consistent (see e.g. \citep[Theorem 13]{yoshida2011polynomial}, \citep[Theorem 1, Lemma 4]{kessler1997estimation}). For the sake of the reader we recall here such results that,  adapted to our case, read:
\begin{theorem}\label{thm:ql-conv}
Let $\tilde H_n := \tilde H_n(\tilde \theta_n, \tilde w_n) = \partial ^2 \ell(\tilde \theta_n, \tilde w_n)$ be the Hessian of $\mathcal \ell_n$ at 
$(\tilde  \theta_n, \tilde  w_n)$. 
	Under assumptions \ref{it:exist} -\ref{it:ident}$'$, 
    \[
    \Gamma_n^{-1}(\tilde \theta_n - \theta_0, \tilde w_n - w_0) \Rightarrow \mathcal N_{\pi_d + d(d-1)}(0, V_{\theta_0}^{-1})
    \]
    and
    \[
    \tilde{\mathcal  H}_n (\theta_0, w_0) \stackrel{p}{\to} V_{\theta_0}, 
    \quad \sup_{|c| \leq \epsilon_n} |  \tilde{\mathcal  H}_n (c + (\theta_0, w_0) ) -  \tilde{\mathcal  H}_n (\theta_0, w_0) | \stackrel{p}{\to} 0 , \quad \epsilon_n \to 0
    \]
\end{theorem}
\noindent where $V_{\theta_0}$ is the positive definite matrix representing the Fisher information of the diffusion. 

\textbf{Adaptive Lasso estimator.} 
We introduce the following loss function 
\begin{gather}
	\mathcal F_n(\theta, w) : = \frac 12 \langle  H_n, (\theta - \tilde \theta_n, w - \tilde w_n)^{\otimes 2}\rangle   + 
	\lambda_n
	\|(\theta, w) \|_{1, \gamma(n,d)}	\notag 
\end{gather}
where $H_n$ is an information matrix,  $\|\cdot \| _{1, \gamma(n,d)}$ denotes the weighted $\ell_1$ norm with 
weight vector $\gamma(n,d) = (\gamma^\alpha_n, 
\gamma^\beta_n, \gamma^w_{n,d})$, i.e.,
\[
\|(\theta, w) \|_{1, \gamma(n,d)} = 
\sum_{i = 1}^{\pi^\alpha} \gamma^\alpha_{n, i} |\alpha_{i}| +
\sum_{i = 1}^{\pi^\beta} \gamma^\beta_{n, i} |\beta_{i}| +
\sum_{1 \leq i, j \leq d, \,i \neq j} \gamma^w_{n,d, ij} |w_{ij}|, 	 
\]
and $\lambda_n > 0$ is a tuning parameter, possibly dependent on the data.
\noindent 
The \emph{adaptive lasso}-type estimator can be formulated as \begin{equation}\label{eq:lasso-est}
	(\hat \theta_n, \hat w_n) \in \arg \min_{\theta, w}  \,\mathcal  F_n(\theta, w).
\end{equation}
Estimator \eqref{eq:lasso-est} allows for simultaneous penalization of the graph-identifying parameters $w$ and of the non-null components in $\theta$.
Denote with $s^\alpha, s^\beta $ the number of non null parameters in 
$\alpha_0, \beta_0$, respectively. 
For ease of notation suppose that the parameter vectors are rearranged so that the first $s^\alpha$ components of $\alpha_0$ are non-null, and similarly for $\beta$. 
The number of non-zero entries in $w$ corresponds to $|E|$.

The analysis of \eqref{eq:lasso-est} depends on the specification of adaptive weights and information matrix 
that satisfy certain assumptions, defined below.  
Let
$\bar \gamma_{n,d}^w = \max_{(i,j) \in E}  \gamma^w_{n,d, ij}$,
$\bar \gamma_{n}^\alpha = \max_{i \leq s^\alpha}  \gamma^\alpha_{n, i}$,
$\bar \gamma_{n}^\beta = \max_{i \leq s^\beta}  \gamma^\beta_{n, i}$ the largest weights for the non-null components, 
and let 
$\check \gamma_{n,d}^w = \min_{(i,j) \notin E}  \gamma^w_{n,d, ij}$,
$\check \gamma_{n}^\alpha = \min_{i \geq s^\alpha}  \gamma^\alpha_{n, i}$,
$\check \gamma_{n}^\beta = \min_{i \geq s^\beta}  \gamma^\beta_{n, i}$
the smallest weight for the null components. 

\begin{enumerate}[label=($\mathbf{L\arabic*}$)]
	
	\item \label{it:ada0} \emph{Consistent information}. 
    The matrix $H_n $ is non-degenerate for $n$ large enough,  and $\mathcal H_n \stackrel{p}{\to} \mathcal H$, where $\mathcal H$ is a positive definite matrix.     
	\item \label{it:ada1} \emph{Adaptive weights} rates of non-null parameters:
	\begin{equation*}
		\frac{ |E| \, \bar \gamma_{n,d}^w}{\sqrt {n\Delta_n}} = O_p(1) ,
		\qquad 
		\frac{ s^\alpha \bar \gamma_{n}^\alpha}{\sqrt n} = O_p(1) ,
		\qquad 
		\frac{ s^\beta \bar \gamma_{n}^\beta}{\sqrt{n\Delta_n}} = O_p(1),
        \qquad 
        \lambda_n = O_p(1).
	\end{equation*}
	\item  \label{it:ada2}
    \emph{Adaptive weights} rates of null parameters:
	\begin{equation*}
		\,\,
		\frac{ \check \gamma_{n,d}^w}{\sqrt{n\Delta_n}}\overset{p}{\longrightarrow}\infty,
		\,\,
		%
	\end{equation*}

\end{enumerate}

Notice that the adaptive coefficients for $w$ may depend on both the sample size $n$ and the dimension $d$.

\begin{remark}
	Condition \ref{it:ada1} depend on the sparsity of the parameter rather than on the full size of the parameter space. Under this framework we can consistently estimate N-SDEs on sparse large graphs.
\end{remark}

A common choice for the adaptive weights is (\cite{zou2006adaptive})
\begin{gather}
	\gamma_{n,j}^\alpha \propto {|\tilde\alpha_{n,j}|^{-\delta_1}},\, j \in [\pi^\alpha],  \quad 
	\gamma_{n,j}^\beta \propto {|\tilde\beta_{n,j}|^{-\delta_2}},\, j \in [\pi^\beta], \\
	\, \gamma_{n, ij}^w \propto  {|\tilde w_{n,ij}|^{-\delta_3}},  (i,j) \in [d]\times [d]
\end{gather}
where $\delta_i>0$. The idea is that the data-driven weights penalize more coefficients whose initial guess has small magnitude.  
The tuning parameter $\lambda_n$ might be chosen by information criteria or validation methods. In \cite{de2024pathwise}, the authors provide algorithms that, for a given sample,  derive a full solution path depending on the tuning parameter. A finite endpoint $\lambda_{\max}$ for such regularization paths can always be found -- that is the smallest tuning parameter such that all the parameters are estimated as zero. Then, one can always choose $\lambda_n$ so that the requirement $\lambda_n = O_p(1)$ is satisfied.
Finally, thanks to \autoref{thm:ql-conv}, the quasi-likelihood based information $\tilde H_n$ provides an example of a converging information matrix. 

\begin{theorem}\label{thm:lasso-cons}
	Under assumptions \ref{it:exist} -\ref{it:ident}$'$ and \ref{it:ada0}-\ref{it:ada1}, the Lasso estimator in \eqref{eq:lasso-est} is consistent, i.e.
	\[
	\Gamma_n^{-1}(\hat \theta_n - \theta_0, \hat w_n - w_0) = O_p(1).
	\]
\end{theorem}

Let $\hat A_n$ the adjacency matrix estimator derived from  \eqref{eq:lasso-est}, i.e.
\begin{equation}\label{eq:adj-est}
	\hat A_{n, ij} = \begin{cases}
		\mathds 1(\hat w_{ij} \neq 0) &  \,i \neq j \\
		0 & i=j
	\end{cases}
	\qquad 1\leq i,j \leq d.
\end{equation}  
Let $\hat G_n= (V, \hat E_n)$ the graph built by means of the estimated adjacency matrix $\hat A_n$.
The next theorem shows the estimated graph coincides with the true graph with probability tending to one.

\begin{theorem}\label{thm:sel-cons}
	Under assumptions \ref{it:exist} -\ref{it:ident}$'$, \ref{it:ada0}, \ref{it:ada1} and \ref{it:ada2}
	\[
	P(\hat G_n = G) \to 1.
	\]
\end{theorem}

\begin{remark}
	\cite{koike2020-glasso} proposed a
	graphical lasso method for the estimation of the covariance matrix 
	of a general semi-martingale setting. Our methodology is different because it allows the estimation of directed graph relations. As an example, see Section~\ref{sec:directed}. We are able to do so because the adjacency matrix appears in the drift equations of the model. 
\end{remark}

After estimating the adjacency matrix one can build a reduced N-SDE model based on the  
estimated neighborhoods $\hat N_i$. 
\begin{align}\label{eq:nsde-2step}
	\de X_t^i &= \bigg(  b_{ii}'(X^i_t, \beta)  + \sum_{j \in \hat N_i}  w_{ij} b_{ij}'(X^i_t, X^j_t; \beta) \bigg) \de t 
	+ 
	\sigma_{i}(X^i_t, \alpha)\de W^i_t 
\end{align}
and re-estimate $\theta = (\alpha, \beta, w_{ij}, \, (i,j) \in \hat E) $ by quasi-likelihood.

\section{N-SDE estimation on synthetic data}\label{sec:sim}
\begin{figure}[h]
	\centering
	\begin{subfigure}[b]{0.3\linewidth}
		\centering
		\includegraphics[width=\linewidth]{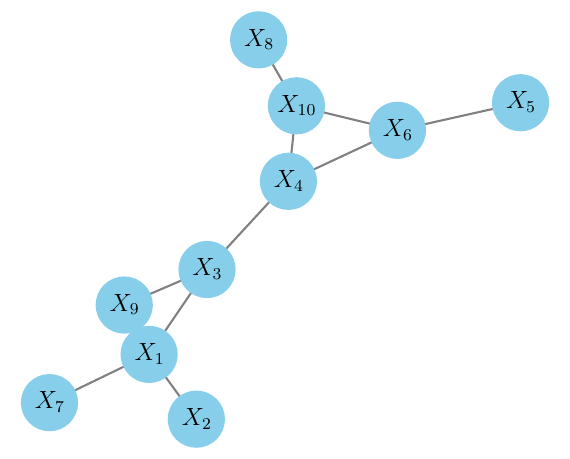}
		\caption{Graph representation of the neighborhood in model \eqref{eq:sde-ex1} for a Erd\H{o}s--R\'enyi random graph}
		\label{fig:graph10a}
	\end{subfigure}
	\hfill 
	\begin{subfigure}[b]{0.6\linewidth}
		\centering
		\includegraphics[width=\linewidth]{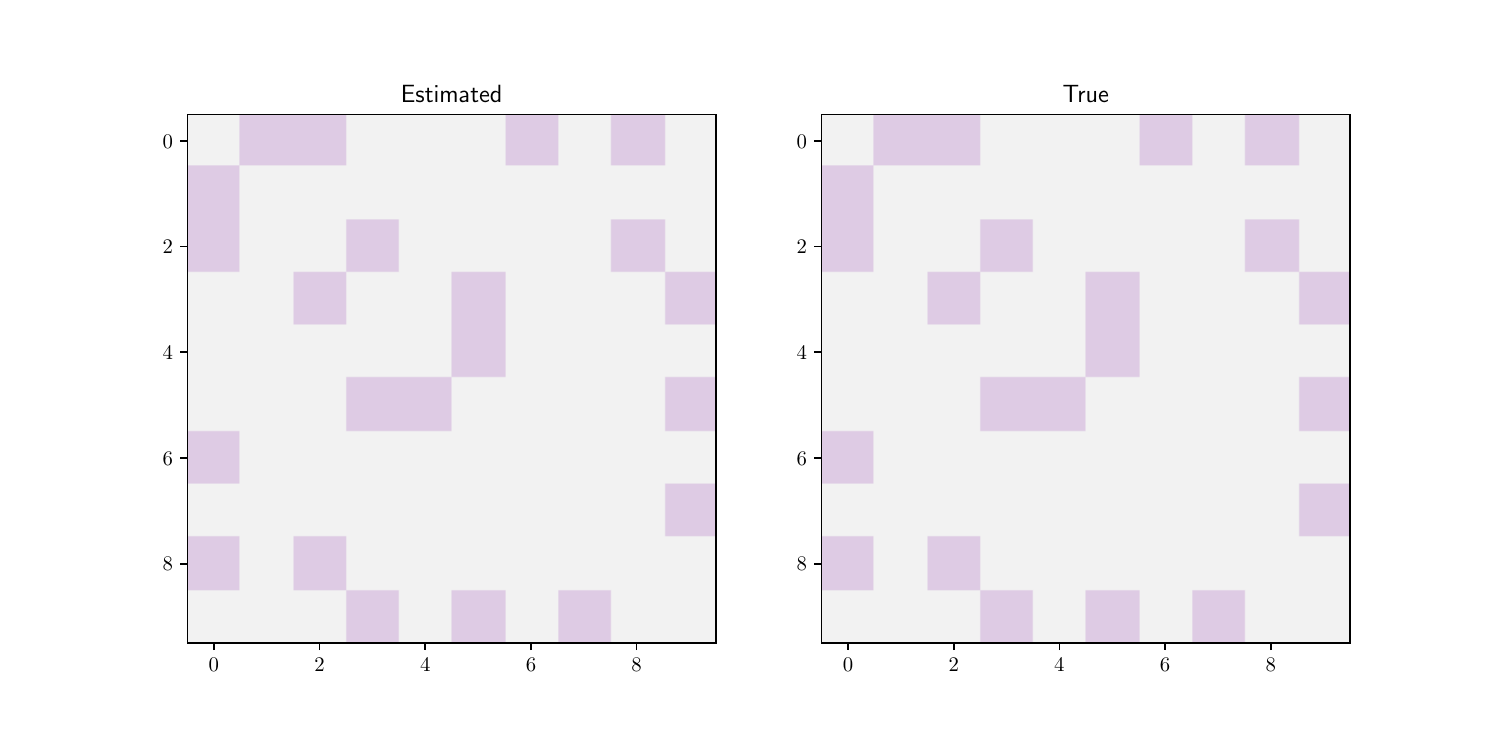}
		\caption{Estimated and true adjacency matrices}
		\label{fig:graph10b}
	\end{subfigure}
	\caption{Erd\H{o}s--R\'enyi random graph}
	\label{fig:graph10}
\end{figure}

\begin{figure}
    \centering
    \includegraphics[width=\linewidth]{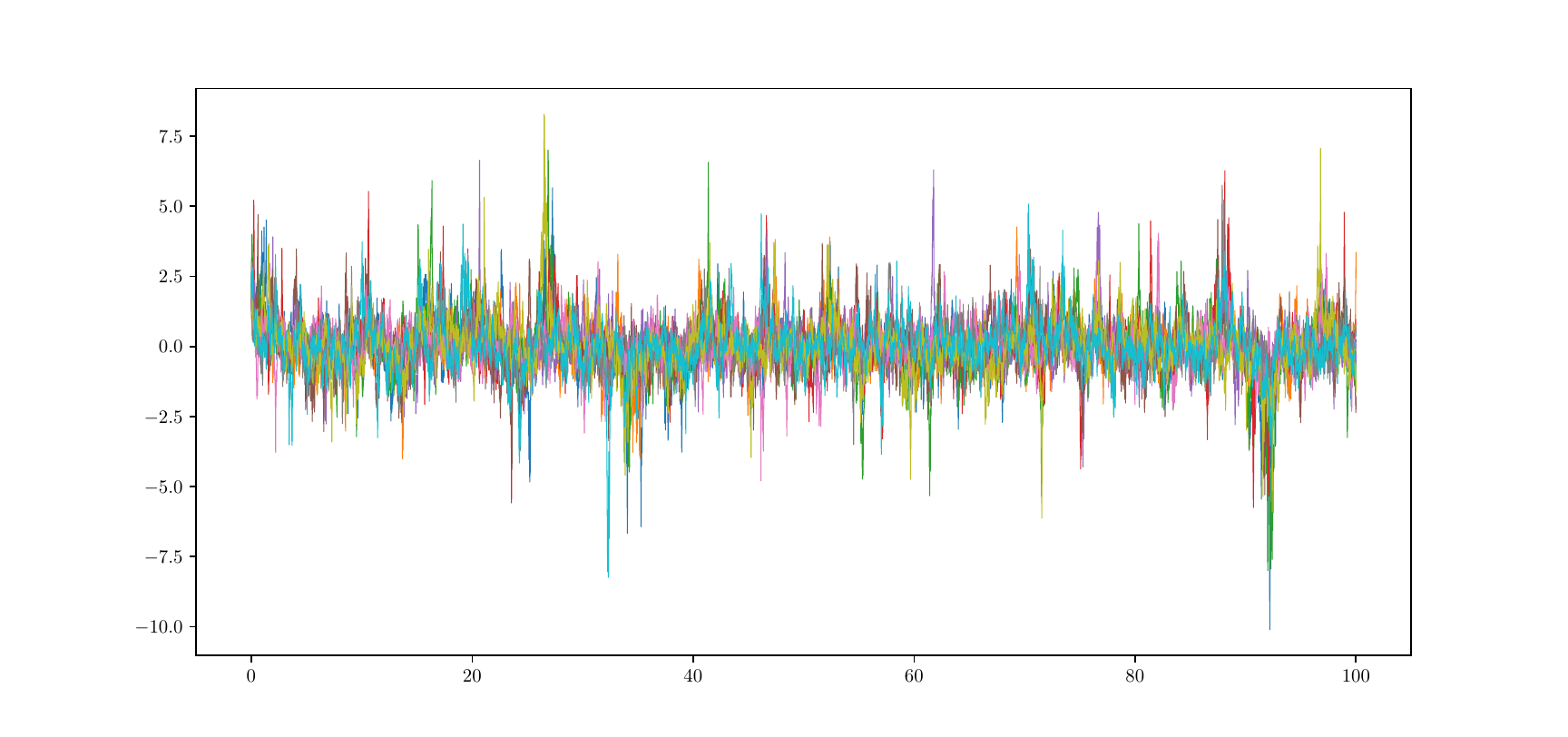}
    \caption{Sample path from model \eqref{eq:sde-ex1}, with true parameter values from \autoref{tab:parameter_estimates}. }
    \label{fig:en-path}
\end{figure}

Consider the following ergodic SDE model
\begin{equation}\label{eq:sde-ex1}
	\de X^i_t = \left(\mu_i X^i_t - \sum_{j \in N_i} \beta_{ij} X^j_t \right) \de t + \alpha_i \, \mathrm{Sigmoid}\left(\sqrt{ 1 + X^2_t}\right) \de W^i_t \qquad i = 1, \ldots, d.
\end{equation} 
where $\mu_i, \beta_{ij} \in \mathbb R$, $\alpha_i >0$. Here we choose as sigmoid function $c \cdot \tanh(x /c)$, for some large value of $c$. 
This function acts as a smooth clipping of the diffusion values. This ensures that the diffusion term is smooth and bounded, and that condition \eqref{eq:ergo1} in \autoref{thm:ergo} is satisfied. 
In practice, by choosing a large value of $c$ (in our simulations we fixed $c=100$) this hardly makes a numerical difference. 
We test our estimation procedure on different graph configurations. In each case we focus on the capability of the model of recovering some different relevant aspect of the graph. 
\paragraph{Erd\H{o}s--R\'enyi graph.}	In this case we are interested in estimating both the parameter values and the graph structure. Here we consider $d = 10$ and the graph is represented in \autoref{fig:graph10a}.
Condition \eqref{eq:ergo-lin} is verified 
as $\tau_{\min}(B) = 5.26$ and $\min \mu_i = 7$. Note also that in this case condition 
$\eqref{eq:ergo-lin2}$ does not hold, as
$8 = \beta_0 \max_i \mathsf{deg}(i) > \mu_0 = 7$, in the notation of Example \ref{ex:lin}. A sample path for this model is shown in \autoref{fig:en-path}.
We set the parameter space to be $[-10^3, 10^3]$ for the real
valued parameters and $[0, 10^3]$ for the non negative parameters and we use the tuning parameter $\delta=1$ 
for the adaptive weights
and
$\lambda = 0.1 \cdot \lambda_{\max}$. 
In order to estimate the graph, we start with a fully connected system, and, since the model is linear, condition (A1') entails
$w_{ij} \equiv \beta_{ij}, \, i,j \in [d]$.
The estimates of the initial quasi likelihood estimator \eqref{eq:ql-est} and the lasso estimator 
\eqref{eq:lasso-est}
are reported in \autoref{tab:parameter_estimates}. We see in \autoref{fig:graph10b} that the adjacency matrix estimator \eqref{eq:adj-est} can recover the graph exactly.

\begin{figure}[h]
	\centering
	\begin{subfigure}[b]{0.30\linewidth}
		\centering
		\includegraphics[width=0.9\linewidth]{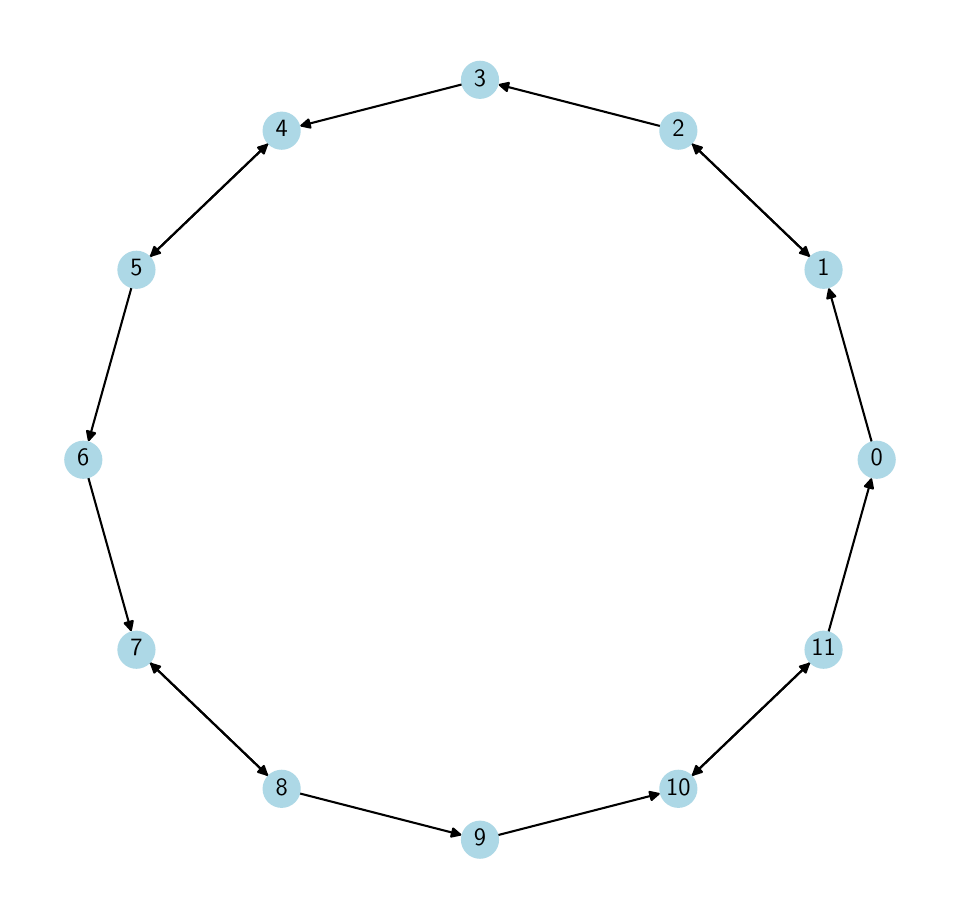}
		\caption{Graph representation of the neighborhood in model \eqref{eq:sde-ex1} for a directed  polymer graph}
		\label{fig:poly12a}
	\end{subfigure}
	\hfill 
	\begin{subfigure}[b]{0.55\linewidth}
		\centering
		\includegraphics[width=\linewidth]{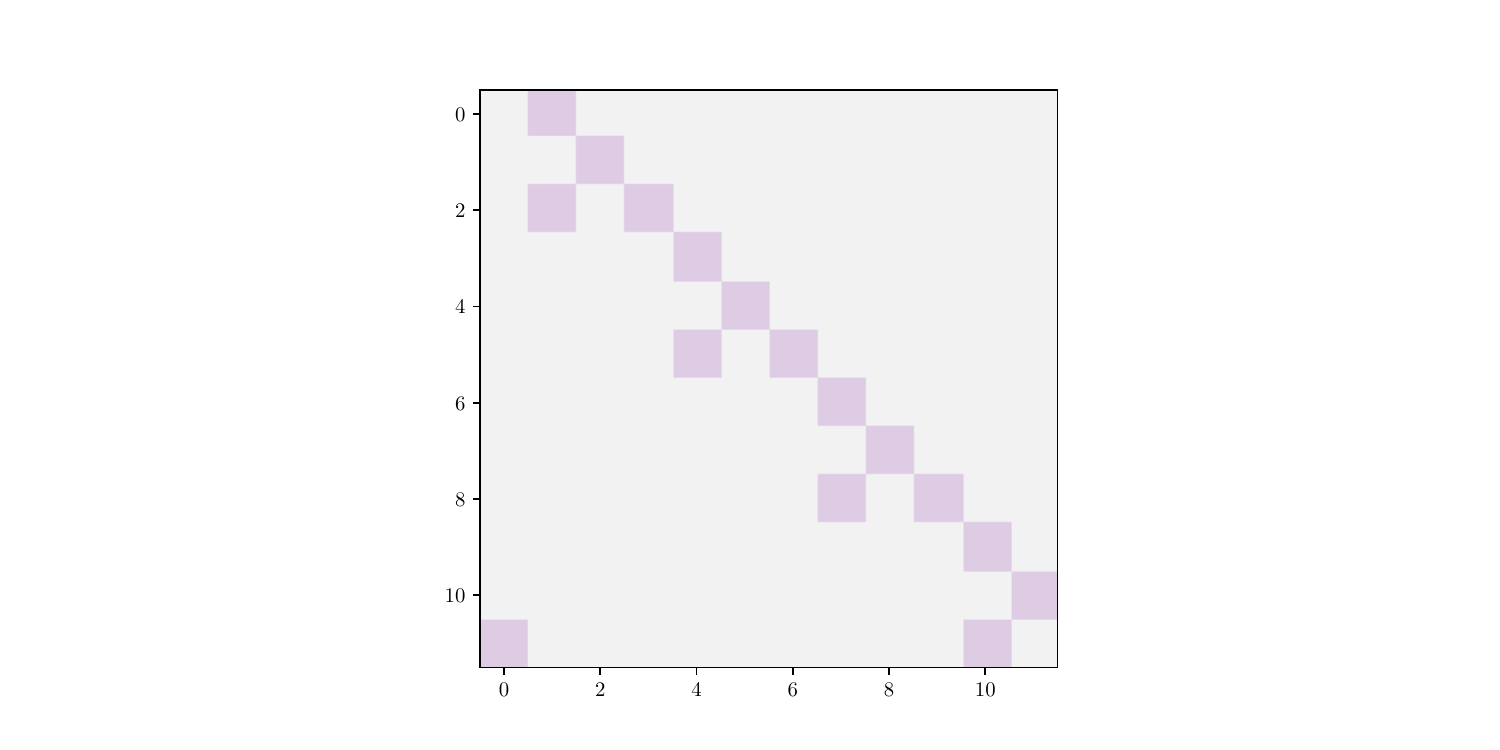}
		\caption{Adjacency matrices}
		\label{fig:poly12b}
	\end{subfigure}
	\caption{Polymer graph}
	\label{fig:poly12}
\end{figure}

\paragraph{Polymer configuration.}\label{sec:directed} In this case we consider a polymer type of graph, $d=12$, following an example in \cite{ou-graph2}.
Here we introduce an important modification, i.e. the graph is oriented. All of the nodes are linked in a chain, but some nodes have  double links, one per direction.  The adjacency matrix is thus not symmetric. The adjacency matrix of the graph is estimated as in \eqref{eq:adj-est}.
The tuning parameter is chosen by evaluating the validation loss, and then by using the more conservative choice $\lambda_{.5 se}$, which corresponds to the minimum of the validation loss plus half its standard deviation.  The graph and the estimated 
adjacency matrix are represented in \autoref{fig:poly12}.
In this case we were able to correctly identify existing relations between nodes as well as the \emph{direction} of such relations.

\begin{figure}[h]
	\centering
	\begin{subfigure}[b]{0.70\linewidth}
		\centering
		\includegraphics[width=0.9\linewidth]{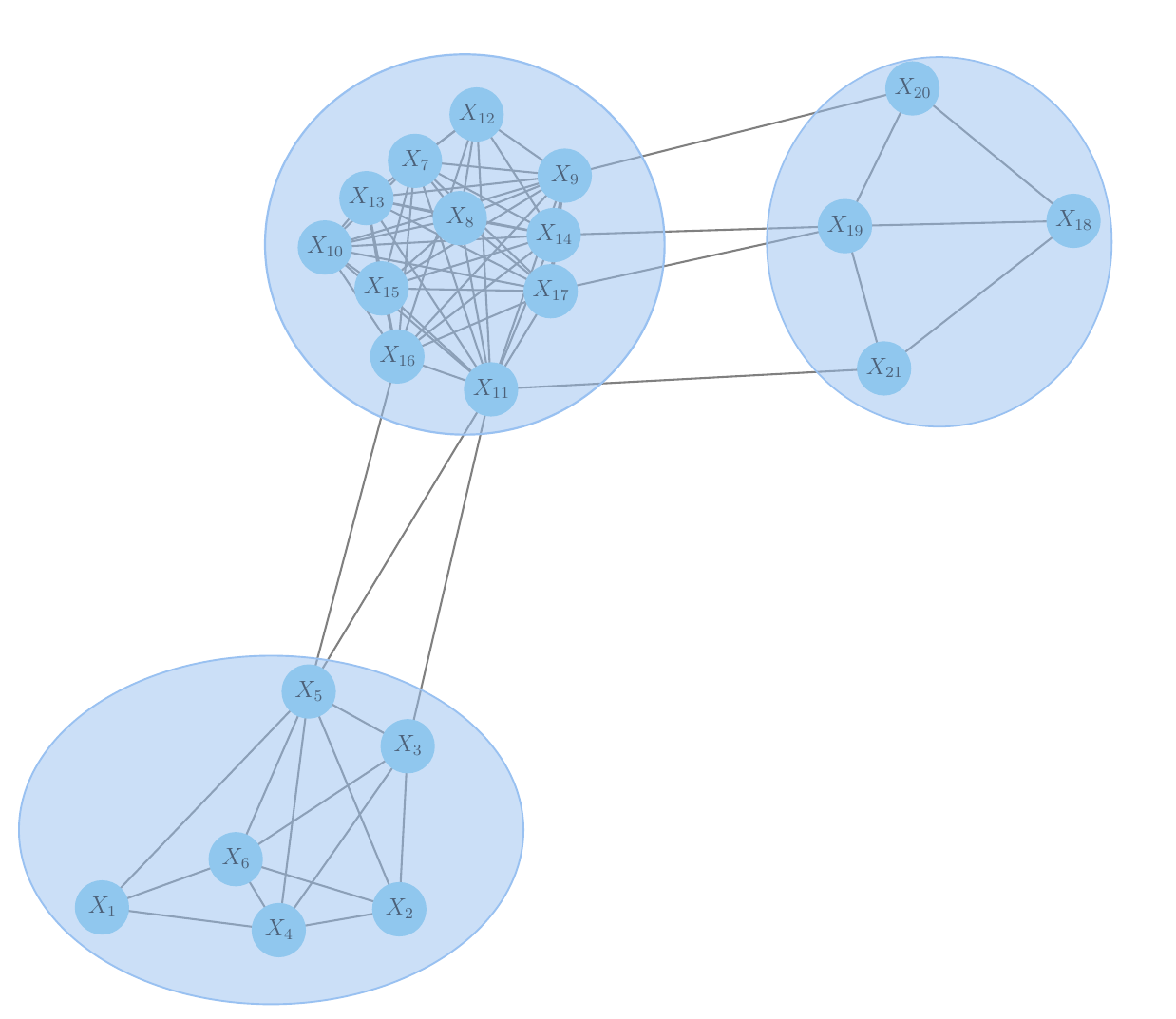}
		\caption{Graph representation of the neighborhood in model \eqref{eq:sde-ex1}  for a stochastic block model graph}
		\label{fig:sbm21a}
	\end{subfigure}
	\hfill 
	
	\begin{subfigure}[b]{0.9\linewidth}
		\centering
		\includegraphics[width=\linewidth]{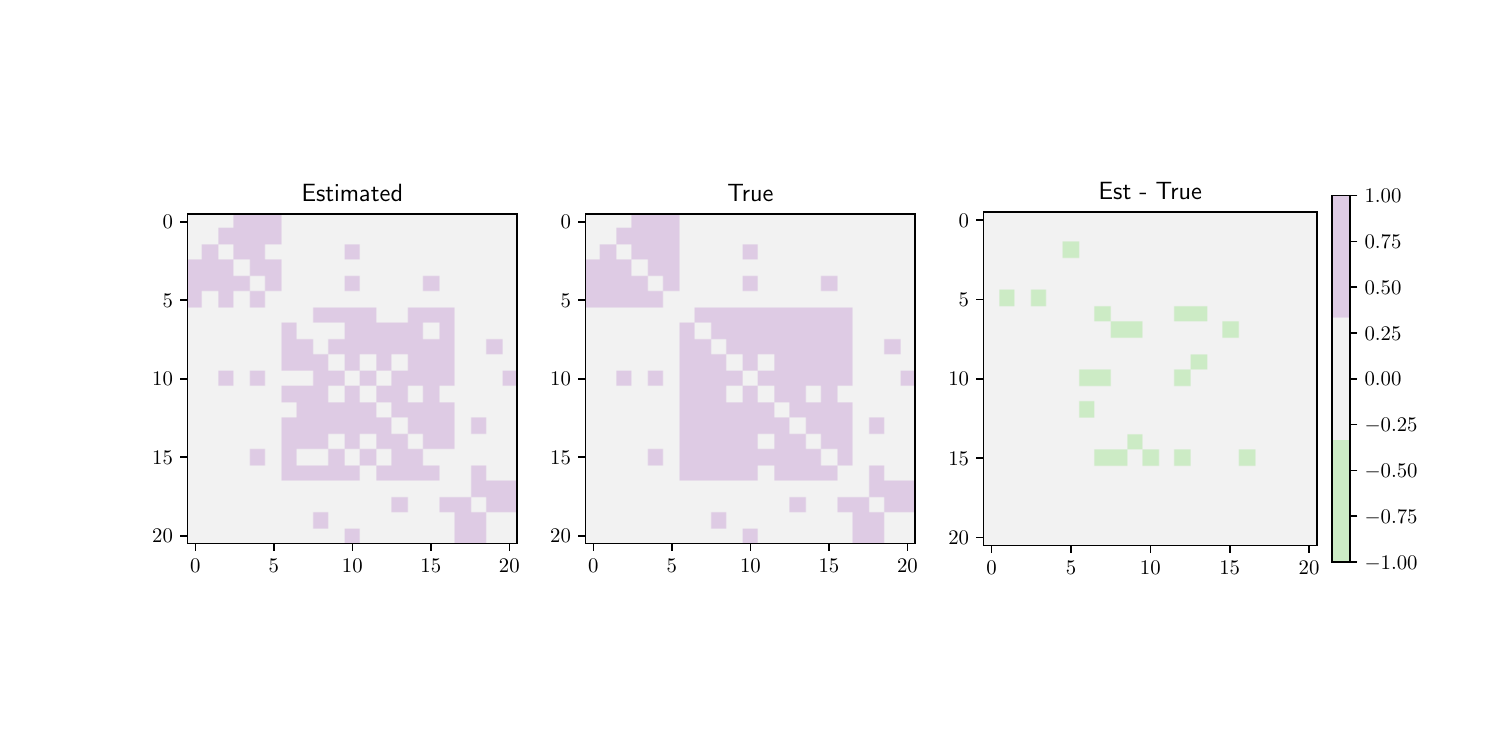}
		\caption{Estimated and true adjacency matrices}
		\label{fig:sbm21b}
	\end{subfigure}
	\caption{Cluster identification in a stochastic block model}
	\label{fig:sbm21}
\end{figure}

\begin{figure}[h]
	\centering
	\includegraphics[width=0.9\linewidth]{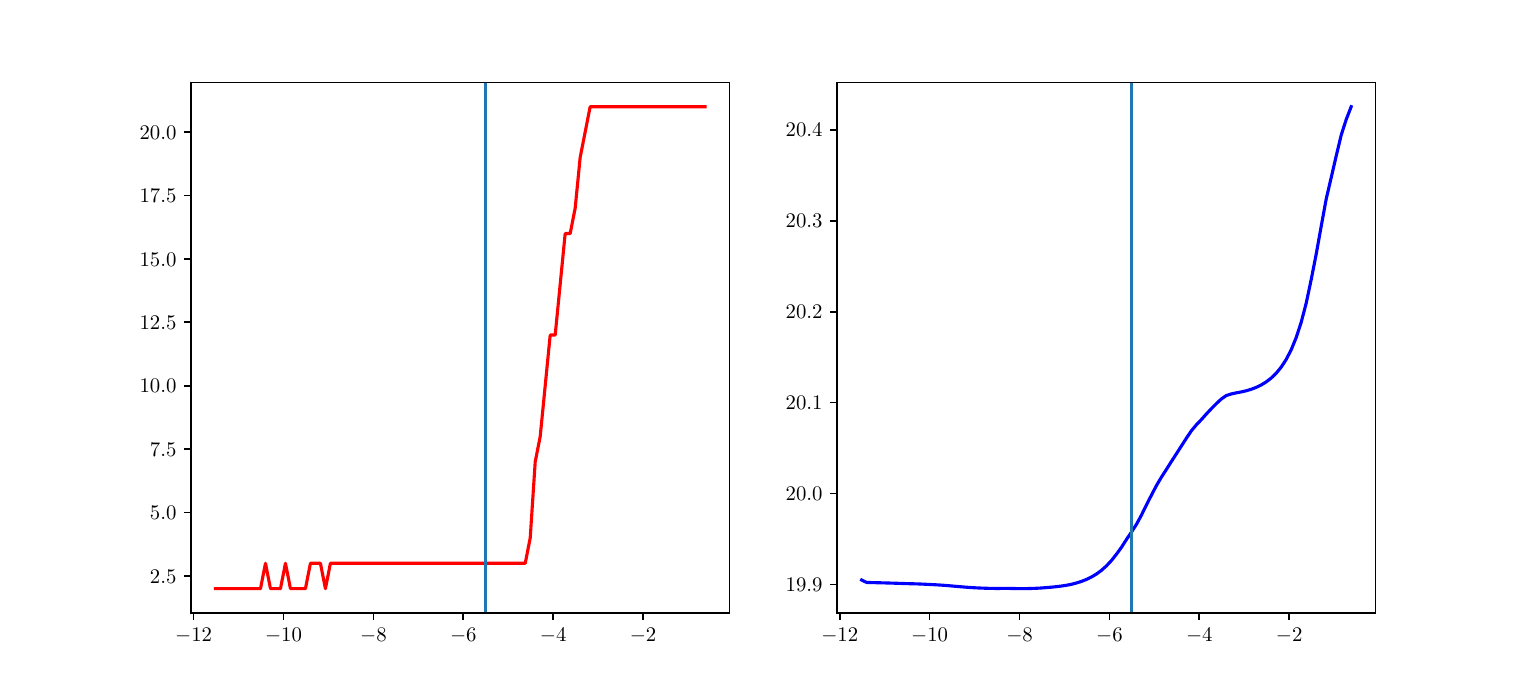}
	\caption{Clusters selected (red) and loss (blue) in a SBM as a function of the penalization parameter, on log-scale. Vertical line represents $\lambda_{0.5 se}$}
	\label{fig:sbm21-clust-sel}
\end{figure}

\paragraph{Stochastic block model.} In this study we aim at recovering the cluster structure of a graph. In order to test this, we consider a graph generated from a stochastic block model. Here $d = 21$, the true graph is made up of three blocks of 4, 11 and 6 nodes respectively with intra-cluster connection probability $p_{in} = 0.9$ and extra-cluster connection probability $p_{ex} = 0.05$. 
The true graph and the communities are shown in \autoref{fig:sbm21a}. 
We first estimate the graph adjacency according to \eqref{eq:adj-est} and then use Louvain community detection algorithm to 
identify the clusters. The edges have been estimated by setting the penalization parameter to  $\lambda_{.5 se}$. The true and estimated adjacency matrices are shown in \autoref{fig:sbm21b}. 
We see that, even though the reconstructed edges do not match perfectly the true ones, our model is capable of identifying the correct cluster structure. 
We also show in \autoref{fig:sbm21-clust-sel} the number of cluster identified as a function of the penalization parameter, compared with the validation loss.

\subsection{Empirical analysis of the error bound}
We now empirically validate the results of \autoref{thm:err-bound}.
In particular, \autoref{tab:simulation_results} contains 
a numerical computation of the mean squared error of the estimator \eqref{eq:lin-est}, for different values of the number of edges, parameters and observation time. Empirical results show perfect agreement with the theoretical bound in \eqref{eq:err-bound} in terms of the expected behaviour as a function of the ratios $K$ and $\epsilon$.  

\begin{table}[h!t]
	\centering
	\begin{tabular}{cccccccccc}
		\toprule
		$d$ & $|E_d|$ & $\pi_d$ & $T$ & $K$ & $\epsilon$ & Bound & Mean Error \\
		\midrule
		8  & 20 & 36 & 10   & 1.8 & 2      & 3.6   & 0.94 (0.34) \\
		&    &    & 20   &     & 1      & 1.8   & 0.52 (0.16) \\
		&    &    & 40   &     & 0.5    & 0.9   & 0.28 (0.08) \\
		&    &    & 80   &     & 0.25   & 0.45  & 0.15 (0.04) \\
		&    &    & 100  &     & 0.2    & 0.36  & 0.12 (0.04) \\
		&    &    & 160  &     & 0.125  & 0.225 & 0.08 (0.02) \\
		&    &    & 200  &     & 0.1    & 0.18  & 0.06 (0.02) \\
		&    &    & 2000 &     & 0.01   & 0.018 & 0.007 (0.002) \\
		\midrule
		16 & 48 & 80 & 96   & 1.7 & 0.5    & 1.35  & 0.23 (0.03) \\
		&    &    & 200  &     & 0.24   & 0.4   & 0.123 (0.02) \\
		\midrule
		32 & 100 & 204 & 200 & 1.45 & 0.5   & 1.2   & 0.23 (0.025) \\
		\bottomrule
	\end{tabular}
	\caption{Simulation Results with different graph configurations for the estimator \eqref{eq:lin-est}}
	\label{tab:simulation_results}
\end{table}

\begin{table}[b]
  \centering
  \tiny
  \setlength\extrarowheight{-10pt}
  \begin{minipage}[b]{0.48\textwidth}
    \centering
    \begin{tabular}{lccc}
      \toprule
      Par. & LASSO & Quasi Lik. & True \\
      \midrule
      $\mu_0$ & 6.3281 & 7.3767 & 7.0 \\
      $\beta_{01}$ & 2.2629 & 2.7986 & 2.0 \\
      $\beta_{02}$ & 1.2856 & 1.6315 & 2.0 \\
      $\beta_{03}$ & 0.0000 & 0.2230 & 0.0 \\
      $\beta_{04}$ & 0.0000 & -0.1383 & 0.0 \\
      $\beta_{05}$ & 0.0000 & 0.4398 & 0.0 \\
      $\beta_{06}$ & 1.7217 & 2.2328 & 2.0 \\
      $\beta_{07}$ & -0.0000 & -0.1943 & 0.0 \\
      $\beta_{08}$ & 1.3903 & 1.7978 & 2.0 \\
      $\beta_{09}$ & 0.0000 & -0.0503 & 0.0 \\
      $\beta_{10}$ & 1.8415 & 2.2509 & 2.0 \\
      $\mu_1$ & 6.6610 & 7.1160 & 7.0 \\
      $\beta_{12}$ & 0.0000 & 0.2183 & 0.0 \\
      $\beta_{13}$ & 0.0000 & -0.1313 & 0.0 \\
      $\beta_{14}$ & -0.0000 & -0.1344 & 0.0 \\
      $\beta_{15}$ & 0.0000 & 0.3437 & 0.0 \\
      $\beta_{16}$ & 0.0000 & 0.0845 & 0.0 \\
      $\beta_{17}$ & 0.0000 & -0.0033 & 0.0 \\
      $\beta_{18}$ & -0.0000 & -0.3980 & 0.0 \\
      $\beta_{19}$ & 0.0000 & 0.1326 & 0.0 \\
      $\beta_{20}$ & 2.2249 & 2.1083 & 2.0 \\
      $\beta_{21}$ & 0.0000 & -0.0236 & 0.0 \\
      $\mu_2$ & 5.6179 & 6.5094 & 7.0 \\
      $\beta_{23}$ & 1.7123 & 2.1529 & 2.0 \\
      $\beta_{24}$ & -0.0000 & -0.1377 & 0.0 \\
      $\beta_{25}$ & 0.0000 & 0.4901 & 0.0 \\
      $\beta_{26}$ & 0.0000 & 0.7233 & 0.0 \\
      $\beta_{27}$ & -0.0000 & -0.6135 & 0.0 \\
      $\beta_{28}$ & 0.3114 & 1.1218 & 2.0 \\
      $\beta_{29}$ & -0.0000 & -0.1869 & 0.0 \\
      $\beta_{30}$ & 0.0000 & -0.0022 & 0.0 \\
      $\beta_{31}$ & 0.0000 & 0.1684 & 0.0 \\
      $\beta_{32}$ & 1.5701 & 1.7314 & 2.0 \\
      $\mu_3$ & 6.6915 & 7.5344 & 7.0 \\
      $\beta_{34}$ & 0.0000 & 0.1736 & 0.0 \\
      $\beta_{35}$ & 1.2632 & 1.6428 & 2.0 \\
      $\beta_{36}$ & 0.0000 & 0.1043 & 0.0 \\
      $\beta_{37}$ & 0.0000 & -0.1132 & 0.0 \\
      $\beta_{38}$ & 0.0000 & 0.5297 & 0.0 \\
      $\beta_{39}$ & 1.9018 & 2.3250 & 2.0 \\
      $\beta_{40}$ & 0.0000 & 0.0117 & 0.0 \\
      $\beta_{41}$ & 0.0000 & 0.2219 & 0.0 \\
      $\beta_{42}$ & -0.0000 & -0.1595 & 0.0 \\
      $\beta_{43}$ & -0.0000 & -0.3041 & 0.0 \\
      $\mu_4$ & 6.1589 & 6.6838 & 7.0 \\
      $\beta_{45}$ & 1.1491 & 1.8099 & 2.0 \\
      $\beta_{46}$ & -0.0000 & -0.5819 & 0.0 \\
      $\beta_{47}$ & 0.0000 & 0.1810 & 0.0 \\
      $\beta_{48}$ & 0.0000 & 0.4505 & 0.0 \\
      $\beta_{49}$ & 0.0000 & -0.1875 & 0.0 \\
       $\beta_{50}$ & 0.0000 & 0.1026 & 0.0 \\
      $\beta_{51}$ & 0.0000 & -0.0571 & 0.0 \\
      $\beta_{52}$ & 0.0000 & -0.1137 & 0.0 \\
      $\beta_{53}$ & 1.7896 & 2.0898 & 2.0 \\
      $\beta_{54}$ & 1.8648 & 2.3059 & 2.0 \\

      \bottomrule
    \end{tabular}
  \end{minipage}
  \quad
  \begin{minipage}[b]{0.48\textwidth}
    \centering
    \begin{tabular}{lccc}
      \toprule
      Par. & LASSO & Quasi Lik. & True \\
      \midrule
           $\mu_5$ & 6.8603 & 7.6419 & 7.0 \\
      $\beta_{56}$ & 0.0000 & 0.2290 & 0.0 \\
      $\beta_{57}$ & -0.0000 & -0.4030 & 0.0 \\
      $\beta_{58}$ & 0.0000 & 0.2169 & 0.0 \\
      $\beta_{59}$ & 1.8530 & 2.4224 & 2.0 \\
      $\beta_{60}$ & 1.6973 & 2.0605 & 2.0 \\
      $\beta_{61}$ & 0.0000 & 0.0537 & 0.0 \\
      $\beta_{62}$ & 0.0000 & -0.0138 & 0.0 \\
      $\beta_{63}$ & 0.0000 & 0.1018 & 0.0 \\
      $\beta_{64}$ & 0.0000 & 0.8902 & 0.0 \\
      $\beta_{65}$ & -0.0000 & -0.5331 & 0.0 \\
      $\mu_6$ & 6.6007 & 7.1529 & 7.0 \\
      $\beta_{67}$ & -0.0000 & -0.3759 & 0.0 \\
      $\beta_{68}$ & 0.0000 & -0.1667 & 0.0 \\
      $\beta_{69}$ & 0.0000 & 0.4044 & 0.0 \\
      $\beta_{70}$ & -0.0000 & -0.1931 & 0.0 \\
      $\beta_{71}$ & 0.0000 & 0.3082 & 0.0 \\
      $\beta_{72}$ & -0.0000 & -0.1411 & 0.0 \\
      $\beta_{73}$ & 0.0000 & 0.0181 & 0.0 \\
      $\beta_{74}$ & -0.0000 & -0.1515 & 0.0 \\
      $\beta_{75}$ & -0.0000 & -0.1822 & 0.0 \\
      $\beta_{76}$ & -0.0000 & -0.1910 & 0.0 \\
      $\mu_7$ & 6.7321 & 7.1909 & 7.0 \\
      $\beta_{78}$ & 0.0000 & 0.2928 & 0.0 \\
      $\beta_{79}$ & 1.8121 & 2.2893 & 2.0 \\
      $\beta_{80}$ & 1.7209 & 1.9590 & 2.0 \\
      $\beta_{81}$ & 0.0000 & 0.1541 & 0.0 \\
      $\beta_{82}$ & 1.8932 & 2.0751 & 2.0 \\
      $\beta_{83}$ & 0.0000 & 0.6526 & 0.0 \\
      $\beta_{84}$ & 0.0000 & 0.2472 & 0.0 \\
      $\beta_{85}$ & 0.0000 & 0.0137 & 0.0 \\
      $\beta_{86}$ & 0.0000 & 0.2700 & 0.0 \\
      $\beta_{87}$ & -0.0000 & -0.5509 & 0.0 \\
      $\mu_8$ & 7.2065 & 7.9472 & 7.0 \\
      $\beta_{89}$ & -0.0000 & -0.3995 & 0.0 \\
      $\beta_{90}$ & 0.0000 & 0.3774 & 0.0 \\
      $\beta_{91}$ & 0.0000 & -0.1424 & 0.0 \\
      $\beta_{92}$ & -0.0000 & -0.1157 & 0.0 \\
      $\beta_{93}$ & 1.9663 & 2.4079 & 2.0 \\
      $\beta_{94}$ & 0.0000 & 0.0438 & 0.0 \\
      $\beta_{95}$ & 1.3233 & 1.8057 & 2.0 \\
      $\beta_{96}$ & -0.0000 & -0.5762 & 0.0 \\
      $\beta_{97}$ & 1.5610 & 2.1589 & 2.0 \\
      $\beta_{98}$ & -0.0000 & -0.3702 & 0.0 \\
      $\mu_9$ & 6.4055 & 7.2089 & 7.0 \\
      \midrule
      $\alpha_0$ & - & 2.0257 & 2.0 \\
      $\alpha_1$ & - & 2.0061 & 2.0 \\
      $\alpha_2$ & - & 2.0341 & 2.0 \\
      $\alpha_3$ & - & 2.0053 & 2.0 \\
      $\alpha_4$ & - & 2.0190 & 2.0 \\
      $\alpha_5$ & - & 2.0145 & 2.0 \\
      $\alpha_6$ & - & 2.0184 & 2.0 \\
      $\alpha_7$ & - & 2.0315 & 2.0 \\
      $\alpha_8$ & - & 2.0603 & 2.0 \\
      $\alpha_9$ & - & 2.0016 & 2.0 \\
      \bottomrule
    \end{tabular}
  \end{minipage}
  \caption{N-SDE parameter estimates in an Erd\H{o}s--R'enyi random graph. Here nodes are labled 0, 1, \ldots, $d -1$, and the parameters are indexed accordingly. No regularization is required on the diffusion part in this example.}
  \label{tab:parameter_estimates}
\end{table}

\section{Applications to real data}\label{sec:app}

\begin{figure}
	\centering
	\includegraphics[width=1\linewidth]{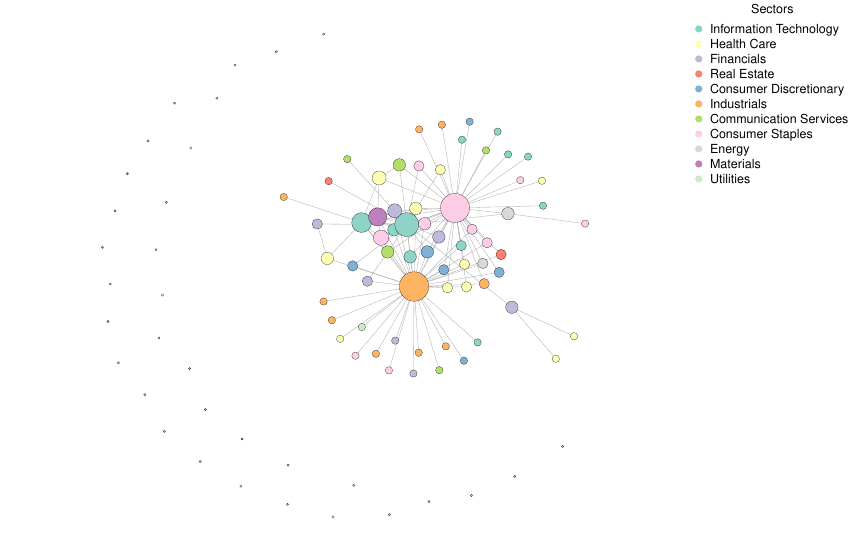}
	\caption{Estimated graph for the components of S\&P100 components stocks.}
	\label{fig:sp-graph}
\end{figure}

In order to test our method on real-world data, we consider high-frequency financial data. 
We take the component stocks of S\&P100 in May 2024. Our observations are closing prices during 5 minutes intervals. Accounting only for complete cases, we have $d=99$ variables  $n=1596$ observations.
We fit estimator \eqref{eq:lasso-est} for a linear drift model \eqref{eq:lin-drift} and constant diffusion. Our enlarged parameter space for $(\theta, w)$ has dimension $\pi_d + d^2=9900$, thus we are in a high-dimensional setting. 
The resulting graph is shown in \autoref{fig:sp-graph}. Vertices are colored according to their Global Industry Classification Standard (GICS) sectors. Our graph exhibits some features 
that have been observed in the literature for similar types of data, see e.g. \cite{koike2020-glasso} and
\cite{barigozzi2018power}.
Our graph consists primarily of a handful of large, connected components with multiple hubs, accompanied by many small isolated components. 
The degree distribution is shown in  . It demonstrates a heavy-tailed pattern, as
the most connected nodes have a disproportionately larger number of links.  
The estimated networks display characteristics that are often observed in power-law graphs (\cite{barigozzi2018power}). 

\begin{figure}
	\centering
	\includegraphics[width=0.6\linewidth]{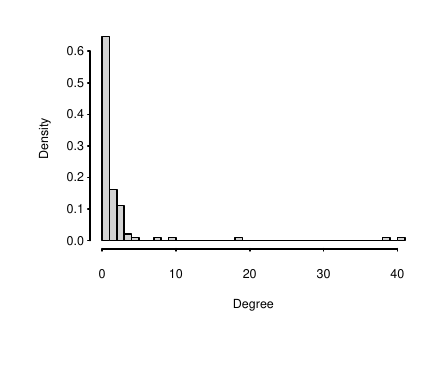}
	\caption{Vertex degree distribution for the S\&P100 graph.}
	\label{fig:enter-label}
\end{figure}

\section{Conclusions.} In this paper we introduce a novel model for stochastic differential equations on networks. This model allows us to deal with high-dimensional systems of SDEs, by modeling the interactions between the series by means of a graph. The novelty in this model lies in the possibility of having general non-linear relations in both the drift interactions and the volatility, as well as directed graph relations. Our contribution is two-fold. On the one hand we provide a form of non-asymptotic control on the estimation error that takes into account the graph scaling in relation to the observation time as well as the graph parametrization. Roughly speaking, this tackles the questions on how much time one needs to observe the graph and how many parameters one can have for each edge in order to have a reliable estimate. On the other hand we analyze a LASSO-based graph estimation procedure, that allows graph recovering  based on the temporal information. We validate our findings by means empirical studies on simulated and real data. 

\section{Proofs}

\begin{proof}[Proof of \autoref{thm:err-bound}.]
	
	We prove that the bound holds true on the event
	$\{|\Gamma_n^{-1}(\hat \theta_n - \theta_0 )|\leq r\}$.
	This event, under assumptions \textbf{A}, has probability at
	least $1 - C_L/r^L$ due to the results in 
	\cite{yoshida2011polynomial}. See, e.g., \cite{yoshida-nonlin} formula (2.14).

	On the event $\{|\Gamma_n^{-1}(\hat \theta_n - \theta_0 )|\leq r\}$, one has that 
	$\{|\hat \theta_n - \theta_0|\leq r/\sqrt{n\Delta_n}\}$
	and so we can apply the inequalities in Assumption $\mathbf C(r/n\Delta_n)$.

	First, by Taylor expansion and Cauchy-Schwartz inequality we have that 
	\begin{align*}
		|\ell_n(\hat \theta_n)  - \ell_n(\theta_0)|
		& \leq 
		\left| \int_0^1 \partial_\theta \ell_n(\theta_0 + u(\hat \theta_n - \theta_0 ) ) \cdot (\hat \theta_n - \theta_0) \mathrm d u \right|
		\\ & \leq 
		\int_0^1 
		|\partial_\theta \overline{\ell_n}(\theta_0 + u(\hat \theta_n - \theta_0 ) )| \mathrm d u 
		\, |\Gamma_n^{-1}(\hat \theta_n - \theta_0)|
		\\
		& \leq 
		\xi_n \sqrt {\pi_d} |\Gamma_n^{-1}(\hat \theta_n - \theta_0)|.
	\end{align*}
	Moreover,
	\begin{align*}
		|\ell_n(\theta_0) - \ell_n(\hat \theta_n)|
		& = 
		\left| \int_0^1 (1-u) 
		\lla \partial^2_{\theta \theta} \ell_n(\hat \theta + u(\theta_0 - \hat \theta_n ) ) , (\hat \theta_n - \theta_0)^{\otimes 2} \rra \mathrm d u \right|     
		\\ & \geq
		\frac{\mu}{2} |\Gamma_n^{-1}(\hat \theta_n - \theta_0)|^2.
	\end{align*}
	
	Putting everything together,
	\begin{align*}
		\sqrt{n \Delta_n} |(\hat \theta_n - \theta_0)| \leq 
		|\Gamma_n^{-1}(\hat \theta_n - \theta_0)| \leq 
		2 \frac{ \xi_n \sqrt {\pi_d}}{\mu}.
	\end{align*}
	By the assumptions \textbf{G}, we then have
	\begin{align*}
		|\hat \theta_n - \theta_0|^2
		&\leq
		4 \frac{ \xi_n^2 {\pi_d}}{\mu^2 n \Delta_n} 
		\leq
		\frac{4 \xi_n^2}{\mu^2} \frac{|G_d| }{n \Delta_n} \frac{\pi_d}{|G_d|} 
		\leq
		\frac{4 \xi_n^2}{\mu^2} K \epsilon .
	\end{align*}
\end{proof}

\begin{proof}[Proof of \autoref{thm:lasso-cons}.]
	We prove consistency by following similar steps as \cite{de2021regularized}, Theorem 1. 
	We begin by writing
	\begin{align*}
		0 &\geq \mathcal F_n(\hat \theta_n, \hat w_n) -\mathcal F_n(\theta_0, w_0) 
		\\ &=
		\frac 12 \langle  H_n, (\hat \theta_n - \theta_0, \hat w_n - w_0)^{\otimes 2}\rangle 
		+ \langle H_n, (\hat \theta_n - \theta_0, \hat w_n - w_0)
		\otimes (\tilde  \theta_n - \theta_0, \tilde  w_n - w_0)\rangle
		\\ & \quad + 
		\lambda_n (\|(\hat \theta_n, \hat w_n) \|_{1, \gamma(n,d)} - 
		\|(\theta_0, w_0) \|_{1, \gamma(n,d)})
		\\ & \geq 
		\frac 12 \|{\mathcal H}_n^{-1}\|^{-1} |\Gamma_n^{-1} (\hat \theta_n - \theta_0, \hat w_n - w_0)|^2
		\\ & \quad - 
		\|{\mathcal H_n}\| |\Gamma_n^{-1} (\hat \theta_n - \theta_0, \hat w_n - w_0)|
		|\Gamma_n^{-1} (\tilde \theta_n - \theta_0, \tilde w_n - w_0)|
		\\ & \quad +
		\lambda_n \left( 
        \sum_{i = 1}^{s^\alpha} \gamma^\alpha_{n, i} (|\hat \alpha_{n, i}| -  |\alpha_{0, i}|) +
  \sum_{i = 1}^{s^\beta} \gamma^\beta_{n, i} (|\hat \beta_{n, i}| -
  |\beta_{0, i}|)+
\sum_{i,j \in E} \gamma^w_{n,d, ij} (|\hat w_{n, ij}| 
 - | w_{0,ij}|)
\right)
\\ & \geq 
  \|{\mathcal H}_n^{-1}\|^{-1} |\Gamma_n^{-1} (\hat \theta_n - \theta_0, \hat w_n - w_0)|^2
		\\ & \quad - 2
		\|{\mathcal H_n}\| |\Gamma_n^{-1} (\hat \theta_n - \theta_0, \hat w_n - w_0)|
		|\Gamma_n^{-1} (\tilde \theta_n - \theta_0, \tilde w_n - w_0)|
		\\ & \quad - 
		\lambda_n \left( \frac{|E|\, \bar \gamma_{n,d}^w}{\sqrt {n\Delta_n}}   +
		\frac{s^\alpha \bar \gamma_{n}^\alpha}{\sqrt{n}}
		+ 
		\frac{s^\beta \bar \gamma_{n}^\beta}{\sqrt{n\Delta_n}}
		\right) |\Gamma_n^{-1} (\hat \theta_n - \theta_0, \hat w_n - w_0)|.
\end{align*}
    
	Hence we get 
	\begin{align*}
		&|\Gamma_n^{-1} (\hat \theta_n - \theta_0, \hat w_n - w_0)| 
		\\ & \quad \leq 
		\|{\mathcal H}_n^{-1}\| \left [  
		2 \|{\mathcal H_n}\| 
		|\Gamma_n^{-1} (\tilde \theta_n - \theta_0, \tilde w_n - w_0)|	
		+
        \lambda_n
		\left( \frac{|E|\, \bar \gamma_{n,d}^w}{\sqrt {n\Delta_n}}   +
		\frac{s^\alpha \bar \gamma_{n}^\alpha}{\sqrt{n}}
		+ 
		\frac{s^\beta \bar \gamma_{n}^\beta}{\sqrt{n\Delta_n}}
		\right) 
		\right]
		\\ &=
		O_p(1) 
	\end{align*}
	because of \autoref{thm:ql-conv} and assumption \ref{it:ada1}.
	
\end{proof}

\begin{proof}[Proof of \autoref{thm:sel-cons}]
	
	%
    The proof is based on a selection consistency and a sign consistency results for adaptive lasso. We split the proof in two steps, that is we prove $(\hat G_n \subset G)$ and $(\hat G_n \subset G)$ with probability tending to 1 (where the inclusion is meant in a non-strict sense). 
    
    \noindent \textit{Step 1.}
    We show that 
    \[
    P(\hat G_n \subset G) \to 1.
    \]
	Denote by $w^\bullet$ the subvector of $w$ corresponding to the null entries of the true parameter $w_0$, that is $w^\bullet = (w_{ij}, (i,j) \notin E)$. 
	This means that $\hat w^\bullet_{n, ij} = 0 \Leftrightarrow (i,j) \notin \hat E_n$, and $(i,j)$ has been correctly excluded.
	Therefore $(\hat w^\bullet_n = 0) \subset (\hat G_n \subset G)$. Then it suffices to show that
	\[
	P(\hat w^\bullet_n \neq 0) \to 0.
	\]
    We follow a standard approach based on the analysis of KKT conditions. 
    Suppose $\hat w_{n, ij} \notin \partial \Theta_w$ and $\hat w_{n, ij} \neq 0$  for some $(i, j) \notin E$. This implies
    \begin{equation}
         \frac{1}{\sqrt{n\Delta_n}} \frac{\partial}{\partial w_{ij}}\mathcal{F}_n(\theta) \Bigg \vert _{\theta=\hat{\theta}} =
         \frac{1}{\sqrt{n\Delta_n}} {H}_n(w_{ij}) (\hat{\theta}_n-\tilde{\theta}_n) + \lambda_n \frac{\gamma^w_{n, ij}}{\sqrt{n\Delta_n}} \mathrm{sgn}(\hat w_{n, ij}) = 0
     \end{equation}
     where $\tilde{H}_n(w_{ij})$  is the row of 
     $\tilde{H}_n$ corresponding to $w_{ij}$.
     Therefore
     \begin{align*}
          \|{\mathcal H}_n(w_{ij})\| 
         \left|\Gamma_n^{-1}(\hat{\theta}_n-\tilde{\theta}_n)\right| & \geq
         \left|
         \frac{1}{\sqrt{n\Delta_n}} {H}_n(w_{ij}) (\hat{\theta}_n-\tilde{\theta}_n)\right|
         \\ & = 
         \left| \lambda_n \frac{\gamma^w_{n, ij}}{\sqrt{n\Delta_n}} \mathrm{sgn}(\hat w_{n, ij}) \right| 
        \geq 
        \lambda_n \frac{ \check \gamma_{n,d}^w}{\sqrt{n\Delta_n}}
     \end{align*}
     where ${\mathcal H}_n(w_{ij})$ denotes the row of the scaled information matrix  is the row of 
     $\tilde{\mathcal H}_n$ corresponding to $w_{ij}$.
     By \autoref{thm:ql-conv}, $\|\tilde{\mathcal H}_n(w_{ij})\|=O_p (1)$; 
     by \autoref{thm:lasso-cons} and \autoref{thm:ql-conv} 
     $\left|\Gamma_n^{-1}(\hat{\theta}_n-\tilde{\theta}_n)\right| = O_p(1)$;
     by \ref{it:ada1} - \ref{it:ada2} 
     $\lambda_n = O_p(1), \,\,  \check \gamma_{n,d}^w/ \sqrt{n\Delta_n} \to \infty$.
     Therefore, for any $i, j \notin E$,
     \begin{align*}
         P\left(\hat{w}_{n, ij}\neq 0, \, \hat{w}_{n, ij}\notin \partial \Theta_w\right)
         & \leq 
         P\left(\|{\mathcal H}_n(w_{ij})\| 
         \left|\Gamma_n^{-1}(\hat{\theta}_n-\tilde{\theta}_n)\right| \geq 
         \lambda_n \frac{ \check \gamma_{n,d}^w}{\sqrt{n\Delta_n}}
         \right) \longrightarrow 0
     \end{align*} as $n\longrightarrow\infty$.
     Moreover, due to the consistency of $\hat \theta_n$, $P(\hat{w}_{n, ij} \in \partial \Theta_w) \to 0$, 
     as $w_0 \in \text{Int}(\Theta_w)$. 
     Therefore
     \[
     P\left(\hat{w}^\bullet_n \neq 0\right)\leq P(\hat w_n \in \partial \Theta_w) 
     + 
     \sum_{(i,j) \notin E}
     P\left(\hat{w}_{n,ij}\neq 0, \, \hat w_{n, ij}\notin \partial \Theta_w\right)\longrightarrow 0.
     \]

    \noindent \textit{Step 2.}
    We show that 
    \[
    P(\hat G \supset G) \to 1.
    \]
	Let $w^\star$ the subvector of $w$ corresponding to non-null 
	entries in $w_0$.
	Note that $\hat w^\star_{n, ij} \neq 0 \Leftrightarrow (i,j) \in \hat E_n$ and $(i,j)$ has been correctly included. 
    Denote by 
    \[
    \mathrm{sgn}(x) = \begin{cases}
        +1 & x > 0 \\
        0 & x=0 \\
        -1 & x < 0.
    \end{cases}
    \]
    Therefore we have the inclusions $
    (\mathrm{sgn} (\hat w^\star_n) =  \mathrm{sgn}(w^ \star_0)  ) \subset (\hat w^\star_{n, ij} \neq  0 \,\,  \forall i,j \in E) \subset (\hat G_n \supset G)$.
	Suppose for simplicity that $\mathrm{sgn}(w_{0,ij}) >0$, for some $i, j \in E$. We have that
	\begin{align*}
		P(\mathrm{sgn} (\hat w^\star_{n, ij}) \neq  \mathrm{sgn}(w_{0,ij}) )
		&= 
		P(\hat w^\star_{n, ij} < 0) = 
		P(\Gamma_n^{-1}(\hat w^\star_{n, ij} - w_{0,ij})  < - \Gamma_n^{-1}  w_{0,ij})
		\to 0
	\end{align*}
	since $\Gamma_n^{-1}(\hat w^\star_j - w_{0,ij}) = O_p(1)$ due to \autoref{thm:lasso-cons}, and $- \Gamma_n^{-1}  w_{0,ij} \to - \infty$. 
	
	Therefore:  
	\[
	P(\hat G_n \not\supset G) \leq 
	 \sum_{i\neq j} P(\mathrm{sgn} (\hat w^\star_{n, ij}) \neq  \mathrm{sgn}(w_{0,ij}) ) \to 0.
	\]

\end{proof}

\printbibliography
	
\end{document}